\documentclass[preprint,10pt]{elsarticle}

\usepackage{amsmath,amssymb,amsfonts}
\usepackage{algorithm, algorithmic}
\usepackage{eurosym}
\usepackage{hyperref}
\usepackage{subcaption}
\usepackage[dvipsnames]{xcolor}
\usepackage{etoolbox}

\DeclareMathOperator*{\argmax}{arg\,max}
\definecolor{forestgreen}{rgb}{0.13, 0.55, 0.13}
\definecolor{cinnabar}{rgb}{0.89, 0.26, 0.2}

\newcommand{\textproc}[1]{{\small\textsf{#1}}}

\newcommand{\negsig}[1]{%
  \ifdimless{#1pt}{.95pt}{#1}{{\color{cinnabar}#1}}%
}

\newcommand{\sig}[1]{%
  \ifdimless{#1pt}{.05pt}{{\color{forestgreen}#1}}{\negsig{#1}}%
}

\newcommand{\fg}[2]{{\color{forestgreen}$#1\cdot10^{#2}$}}

\makeatletter
\def\ps@pprintTitle{%
	\let\@oddhead\@empty
	\let\@evenhead\@empty
	\def\@oddfoot{}%
	\let\@evenfoot\@oddfoot}
\makeatother

\begin{document}

\begin{frontmatter}



\title{Jasmine: a New Active Learning Approach to Combat Cybercrime}


\author[a1]{Jan Klein\corref{cor1}}
\cortext[cor1]{Corresponding author}
\ead{j.g.klein@cwi.nl}
\author[a2]{Sandjai Bhulai}
\ead{s.bhulai@vu.nl}
\author[a3]{Mark Hoogendoorn}
\ead{m.hoogendoorn@vu.nl}
\author[a1]{Rob van der Mei}
\ead{r.d.van.der.mei@cwi.nl}

\affiliation[a1]{organization={Department of Stochastics, Centrum Wiskunde \& Informatica},
            addressline={Science Park 123}, 
            postcode={1098XG},
            city={Amsterdam},
            country={The Netherlands}}
\affiliation[a2]{organization={Department of Mathematics, Vrije Universiteit},
            addressline={De Boelelaan 1111}, 
            postcode={1081HV},
            city={Amsterdam},
            country={The Netherlands}}
\affiliation[a3]{organization={Department of Computer Science, Vrije Universiteit},
            addressline={De Boelelaan 1111}, 
            postcode={1081HV},
            city={Amsterdam},
            country={The Netherlands}}

\begin{abstract}
Over the past decade, the advent of cybercrime has accelarated the research on cybersecurity. However, the deployment of intrusion detection methods falls short. One of the reasons for this is the lack of realistic evaluation datasets, which makes it a challenge to develop techniques and compare them. This is caused by the large amounts of effort it takes for a cyber analyst to classify network connections. This has raised the need for methods \textit{(i)} that can learn from small sets of labeled data, \textit{(ii)} that can make predictions on large sets of unlabeled data, and \textit{(iii)} that request the label of only specially selected unlabeled data instances. Hence, Active Learning (AL) methods are of interest. These approaches choose specific unlabeled instances by a query function that are expected to improve overall classification performance. The resulting query observations are labeled by a human expert and added to the labeled set.

In this paper, we propose a new hybrid AL method called Jasmine. Firstly, it determines how suitable each observation is for querying, i.e., how likely it is to enhance classification. These properties are the uncertainty score and anomaly score. Secondly, Jasmine introduces dynamic updating. This allows the model to adjust the balance between querying uncertain, anomalous and randomly selected observations. To this end, Jasmine is able to learn the best query strategy during the labeling process. This is in contrast to the other AL methods in cybersecurity that all have static, predetermined query functions. We show that dynamic updating, and therefore Jasmine, is able to consistently obtain good and more robust results than querying only uncertainties, only anomalies or a fixed combination of the two. 
\end{abstract}


\begin{highlights}
\item Jasmine is a novel hybrid Active Learning method for Network Intrusion Detection
\item Jasmine queries uncertain, anomalous and randomly selected observations
\item Jasmine can dynamically adjust the balance between these query types
\item Jasmine can learn the best strategy during the labeling process
\item Jasmine is the first Active Learning method with dynamic updating
\item Jasmine performs better than existing static Active Learning methods
\end{highlights}


\end{frontmatter}


\section{Introduction}
\label{sec:intro}

The fight against cybercrime has become a priority for many countries. This is with good reason, because the average cost of a single cyberattack in Europe is around $50$ thousand euros, as estimated by Forrester Consulting and Hiscox in 2020~\cite{consultancy}. Most common attacks were found to be DDoS attacks, malware infections and phishing, and these were mostly targeted on companies and even governmental institutes. For example, 68\% of Dutch firms reported at least one cyber incident in 2019. Therefore, the amount of money that the surveyed European companies invested in cybersecurity has increased by 39\% from $1.3$ million to $1.8$ million euros. In academia, the awareness for research in the field of cybersecurity has also grown. For instance, Mouloua et al.\ analyzed the articles published in the Proceedings of the Human Factors and Ergonomics Society (HFES) from 1980 to 2018~\cite{mouloua2019trend}. They showed that $73\%$ of articles related to cybersecurity published in that almost 40-year span were written in the last nine years. 

Since most cyberattacks are aimed at companies and countries, it is important to know how networks of computers can be protected. A Network Intrusion Detection System (NIDS) is software designed to detect unusual, malicious events in a computer network. There are several types of systems with each having their own set of challenges for which several solutions have been proposed~\cite{sommer2010outside, zamani2013machine, sultana2019survey}. However, there are some broader challenges in cybersecurity research.
Most importantly, Xin et al.\ and Yang et al.\ argue that not much consideration is given to deployment efficiency~\cite{xin2018machine, yang2018active}. This means that not much is practically done with published research, because of time complexity of the techniques and the efficiency of detection in actual networks. The latter is due to the lack of realistic datasets, since it takes a lot of time and effort for a human to classify network connections correctly. 
Moreover, cyber analysts have to label many redundant connections just to construct a representative dataset.
This loads the experts with tedious work and leads to an underuse of their capabilities. Therefore, 
it would be beneficial to only present `informative' network connections to the cyber analyst. This can be realized in Active Learning (AL), in which the model chooses from which unlabeled data instances it wants to learn and then queries their labels~\cite{settles2009active, kumar2020active}.

Several AL methods have been proposed in intrusion detection research. Many of them focus on querying \textit{uncertain} data, i.e., requesting the label of observations about which the model is not sure how to classify them~\cite{li2007active, torres2019active, gornitz2009active}. Adding these observations with their correct label is expected to enhance classification performance more quickly than randomly selecting observations. Other studies consider different query strategies or combine several query approaches to make the AL procedure more robust~\cite{yin2018active}. Stokes et al.\ query both uncertain and \textit{anomalous} instances~\cite{stokes2008aladin}. The latter are observations that behave vastly differently than expected and that could indicate malicious activities. Although combining query types increases prediction performance, the optimal balance of the types depends on, for example, the dataset. Moreover, the balance has to be determined beforehand and is still up for debate.

In our research, we propose a novel AL method called Jasmine that introduces \textit{$\alpha$-dynamic updating}. This allows our method to adjust the balance between querying different types of observations such that the right types are proposed to the human expert at the right time. This makes sense, because the structure of the labeled set (on which the classifier is trained) changes during the labeling procedure. Hence, Jasmine is able to learn the best query strategy during the process. The types of instances that Jasmine considers potentially informative are uncertain, anomalous and randomly selected observations. 
Querying uncertain and anomalous unlabeled data is inspired by the AL method of Stokes et al.\ that queries these observations types in a fixed 50/50 split~\cite{stokes2008aladin}. 
Jasmine is able to dynamically change the balance to ensure that the most informative observations to the intrusion detection model are queried. This sets our method apart from existing AL methods, because (to our knowledge) Jasmine is the only method that allows for this. 
Our contributions are:
\begin{itemize}
	\item We propose a new AL method called Jasmine that introduces dynamic updating of the balance between querying uncertain, anomalous and randomly selected unlabeled data. 
	\item We present the mathematical formulation of Jasmine and explain why it can find a good balance given the current labeling state. 
	\item We apply Jasmine to two commonly used network intrusion detection datasets and use them in different experimental settings. We show that Jasmine obtains good results and is more robust than static query approaches. Therefore, Jasmine is more reliable to use, because it can adapt itself to different situations. 
\end{itemize}

The rest of the paper is organized as follows. In Section~\ref{sec:relatedwork}, we explain the AL paradigm and what methods have been proposed in intrusion detection. Section~\ref{sec:preliminaries} introduces the mathematical notation used in the AL framework. In Section~\ref{sec:methods}, we propose our method Jasmine and explain how it works. The experiments that we execute to validate our method are discussed in Section~\ref{sec:expsetup}. Section~\ref{sec:results} subsequently shows the results of these experiments and their interpretations. Finally, in Section~\ref{sec:discussion}, we draw conclusions about this study and make suggestions for further research.



\section{Related Work}
\label{sec:relatedwork}


\begin{figure}[h!]
 \centering
 \includegraphics[width=7cm]{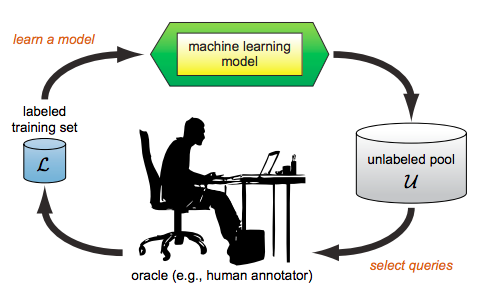}
 \caption{Illustration of Active Learning framework (taken from Settles~\cite{settles2009active}).}
 \label{fig:activelearning}
\end{figure}

Active Learning is a subfield of machine learning in which the premise is that only a small part of the data is labeled, while the labels for the rest of the data are not specified. The AL procedure is illustrated in Figure~\ref{fig:activelearning}. Firstly, an ML model is trained on the labeled set and applied to the unlabeled observations to obtain output predictions. Then, interesting observations from the unlabeled part are selected and an oracle is asked to provide the actual labels. Subsequently, these query observations are added to the labeled set and the procedure continues.  An important advantage of AL is that the model needs less train data to learn better~\cite{settles2009active, kumar2020active}. This is especially beneficial in domains in which it is laborious to label data. Examples of such fields are speech recognition, information extraction (person names from texts, annotation of genes, etc.) and classification of files (relevant or irrelevant documents, images, etc.). Network intrusion detection is one of the fields in which it is burdensome to label the huge number of network connections.

\subsection{Query Strategies}

There are multiple ways to query the oracle, but we focus on \textit{pool-based sampling}. Here, specific observations are selected to be queried from the unlabeled pool. The decision is based on some informativeness measure, which is determined for all instances in the pool (or a subset thereof). This approach to querying has been applied to many real-world problems, including the fields mentioned in the previous paragraph. It works well with a human oracle and when it is relatively easy to compute the informativeness of all observations at once. 

As we mentioned before, a commonly used informativeness measure is the uncertainty score. This measure is used to query observations about which the model is the least certain on how to predict their label~\cite{lewis1994sequential}. This is a simple informativeness measure, because no new models have to be trained and only the output of the classifier is required to determine the uncertainty of each observation.

\subsection{Active Learning in Network Intrusion Detection}

Which specific query strategy, informativeness measure(s) and ML techniques are used depends on the AL application. In network intrusion detection, several AL methods have been proposed. These approaches mostly rely on uncertainty sampling. Li et al.\ use Transductive Confidence Machines for K-Nearest Neighbors for supervised intrusion detection with uncertainty sampling and query by committee~\cite{li2007active}. 
Guerra Torres et al.\ make use of Random Forests for prediction and query uncertain observations.~\cite{torres2019active}
G\"{o}rnitz et al.\ use a Support Vector Domain Description (SVDD) for anomaly detection with uncertainty sampling as the AL component~\cite{gornitz2009active}. 
All these studies show that a method with AL obtains better results than one without it, or that the proposed query strategy performs better than randomly presenting observations to the oracle. However, they are rather limited in the sense that they only consider one informativeness measure for query selection.

Though, there are studies that incorporate two measures to improve classification performance. Yin et al.\ use an SVDD~ for prediction, and combine the uncertainty informativeness measure with density information as the query approach. More specifically, the distance of an observation to its nearest neighbor is calculated and instances residing in high density areas are more likely to be queried~\cite{yin2018active}.
Stokes et al.\ propose to combine uncertainty sampling with querying anomalous data, thus presenting instances that behave vastly differently than expected to the oracle~\cite{stokes2008aladin}. 
These studies show that using multiple informativeness measures further improves performance, because more characteristics of the data are used to determine which unlabeled observations would improve predictions the most if they were labeled. 

The common factor of the aforementioned research is that all proposed query functions are \emph{static} in nature. From the start, it is exactly known how the set of query observations is constructed and this approach cannot be changed. This means the contribution of each informativeness measure in the selection of the query observations has to be fixed beforehand. However, the optimal balance of these contributions depends on the overall structure of the data and also on the current state of the labeling process. Therefore, we consider a \emph{dynamic} query approach in this research to address these problems. To this end, the balance can be adapted during the procedure to best fit the data. More specifically, the method learns the distribution of query types that is expected to increase prediction performance the most given the current state.


\subsection{ALADIN}

The AL methodology for network intrusion detection that we use as a starting point for our research was developed by Stokes et al.\ and is called ALADIN~\cite{stokes2008aladin}. 
It was chosen because it combines two important informativeness measures: the \textit{(i)} uncertainty score and \textit{(ii)} anomaly score. 
On the one hand, by querying selected anomalies, new classes of network traffic can be found within the data. On the other hand, by querying uncertainties, the accuracy of the classifier should increase in the next time step when the correct classes have been provided by the human expert.
\begin{figure}[h!]
 \centering
 \includegraphics[width=6.5cm]{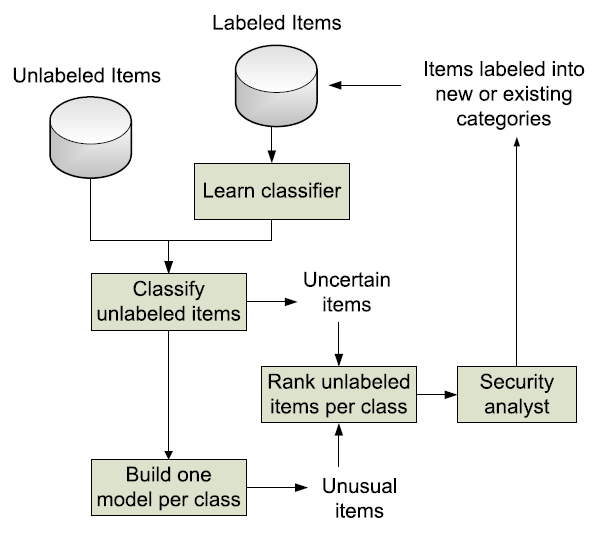}
 \caption{Schematic representation of an ALADIN iteration. Taken from Stokes et al.~\cite{stokes2008aladin}.}
 \label{fig:aladin}
\end{figure}
ALADIN combines both Pelleg's algorithm for anomaly detection~\cite{pelleg2004active} and Almgren's algorithm for intrusion detection~\cite{almgren2004using}. In the first algorithm, an anomaly detection model is constructed for each data class, so one for the benign class and one for the malicious class in the binary case. 
How unlikely an unlabeled observation is according to the model of its (predicted) class, determines how anomalous it is. The less likely an instance is, the more interesting it is for querying. In the second algorithm, observations that lie close to the decision boundary of the trained classifier are deemed uncertain, and hence, interesting for querying. 
The description of ALADIN is illustrated in Figure~\ref{fig:aladin}.

On the 1999 KDD-Cup dataset~\cite{stolfo1999kdd}, Stokes et al.\ show that ALADIN achieved high accuracy in predicting known traffic classes. Moreover, it was able to detect previously unknown categories quickly. The authors also applied their method on real corporate network logs containing about 13 million observations. Interestingly, ALADIN discovered a new trojan, which was not found by the rule-based NIDS that the company used. This shows that incorporating anomalous and uncertain observations in the query set led to favorable results. 

ALADIN uses simple machine learning techniques such as logistic regression and naive Bayes to be able to scale well. However, the authors mention that the benign class in the KDD-Cup dataset is very diverse, meaning that this class may not be easily predicted with the logistic regression classifier. In our research, we consider more advanced machine learning techniques that are better able to capture the diversity of network traffic. More importantly, in ALADIN, half of the queried observations are anomalous, while the other half is uncertain. This fixed 50/50 split is not motivated by the authors. Hence, model performance can improve when the proportion between querying anomalies and uncertainties is changed depending on the considered dataset and within the process. This leads to our $\alpha$-dynamic updating. 


\section{Preliminaries}
\label{sec:preliminaries}

Before we provide the mathematical formulation of Jasmine, we introduce some notation to describe the AL framework in general. Firstly, we assume to have a dataset $\mathbf{X} \in \mathbb{R}^{M \times K}$, with $M \in \mathbb{N}$ the number of observations and $K \in \mathbb{N}$ the number of features or attributes. Let $\mathbf{x}_i \in \mathbb{R}^K$ be the feature vector of observation $i \in \{1, \dots, M\}$ and let $y_i \in \{0, 1, *\}$ be its corresponding response value. Here, `0' represents the benign class and `1' the malicious class. The symbol `*' means that the class or label is missing. In AL, it is assumed that only a (possibly small) part of the data is labeled, while the rest is unlabeled. 
The set of labeled observations $\mathcal{L}(t)$ and the set of unlabeled observations $\mathcal{U}(t)$ depend on the iteration or time step $t = 1, \dots, T$, where $T \in \mathbb{N}$ is a predetermined maximum number of iterations. 
Since more labels become available when the iteration procedure progresses, the vector of response values of all the instances $\mathbf{y}(t) = (y_1(t), \dots, y_M(t))$ is dependent on time. Now, for all iterations it holds that $\mathcal{L}(t)$ and $\mathcal{U}(t)$ are disjoint and their union is the complete dataset $(\mathbf{X}, \mathbf{y}(t))$ with labels up to time $t$. Let $L(t) := |\mathcal{L}(t)|$ be the number of labeled observations and $U(t) := |\mathcal{U}(t)|$ the number of unlabeled instances at time $t$. Note that $L(t+1) > L(t)$, while $U(t+1) < U(t)$, because every iteration the human expert adds labels to previously unlabeled observations. $\mathcal{L}(0)$ contains the instances that are labeled from the start, while $\mathcal{U}(0)$ consists of all initially unlabeled observations. In iteration $t$, a supervised machine learning technique $f_t: \mathbb{R}^K \rightarrow [0,1]$ is trained on $\mathcal{L}(t-1)$. This classifier $f_t$ is then applied to $\mathcal{U}(t-1)$ to obtain predictions for the actual classes of the unlabeled observations. Then, a query function $\psi$ determines which instances from $\mathcal{U}(t-1)$ are selected to be queried to the human expert. Let $\mathcal{Q}(t) \subseteq \mathcal{U}(t-1)$ be the constructed set of query observations. Usually, this set has fixed size $Q := |\mathcal{Q}(t)|$. Then, the expert provides the labels $y_q(t) \in \{0,1\}$ for the observations $q \in \mathcal{Q}(t)$. Note that in the previous iteration their labels were still unknown: $y_q(t-1) = \text{`*'}$.

\section{Methods}
\label{sec:methods}

In this section, we introduce our Active Learning method Jasmine. Its key component is $\alpha$-dynamic updating, which allows the model to dynamically adjust the balance between querying anomalous, uncertain and random observations during the procedure. The adjustment of the balance comes in two flavors. Firstly, the initial proportions of the three types of query observations are determined. Based on the available initially labeled data $\mathcal{L}(0)$, it can be beneficial to start with querying 60\% anomalous, 15\% uncertain and 25\% random observations, for example. Secondly, the balance between the types can be changed during the labeling process, because it may be better to query increasingly more anomalous instances when more and more labels are provided by the oracle. This leads to dynamically updating the query fractions. 
Moreover, we also consider querying random observations. This may seem counter-intuitive in the AL setting, because its premise is not to bother the human expert with labeling redundant observations. However, when $\mathcal{L}(0)$ is not a good representation of the entire dataset, querying some random observations could be beneficial. 

\begin{figure}[h!]
 \centering
 \includegraphics[height=8cm]{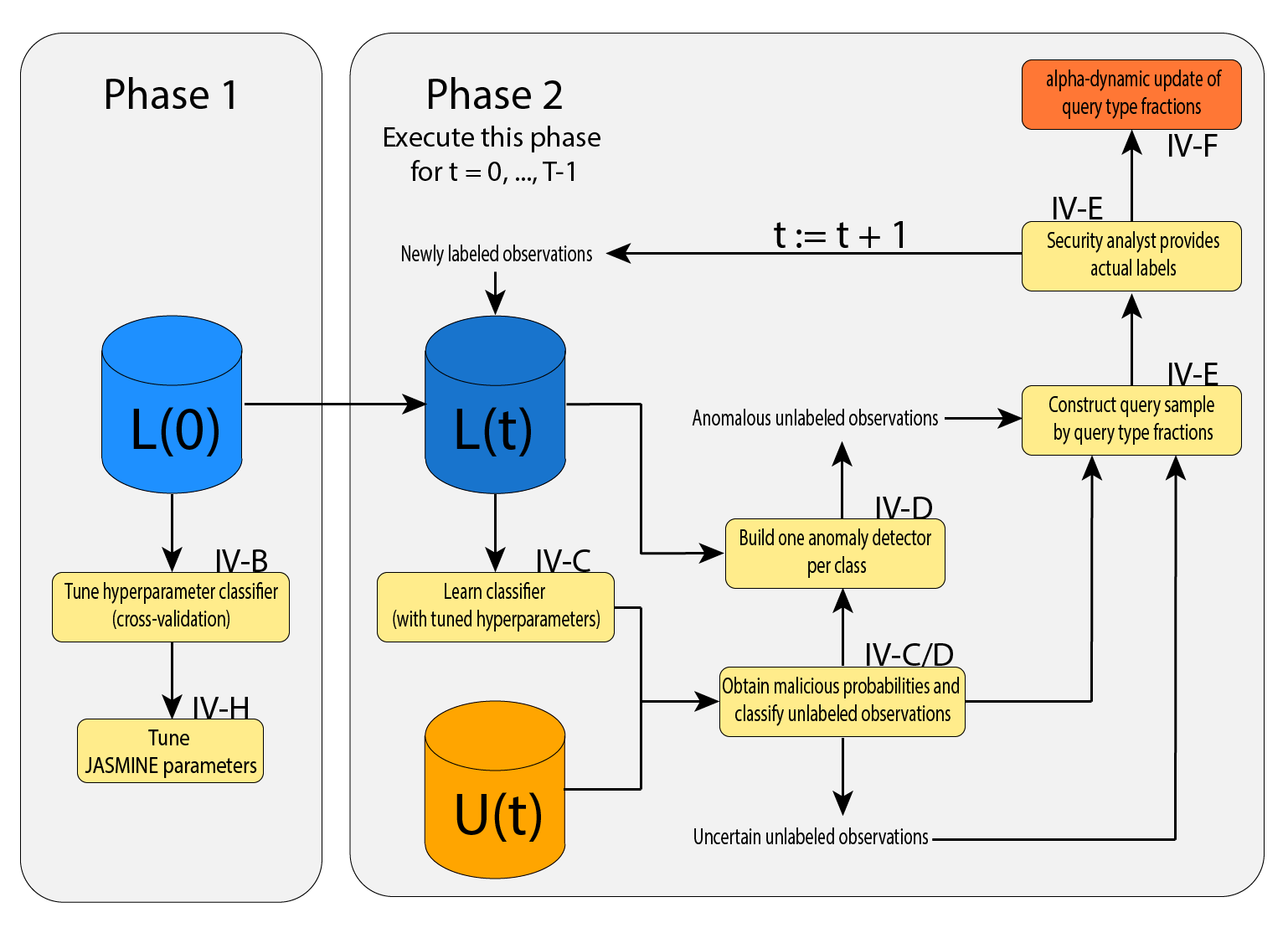}
 \caption{Schematic representation of complete Jasmine procedure.}
 \label{fig:jasmine}
\end{figure}
The complete Jasmine procedure is illustrated in Figure~\ref{fig:jasmine}. It consists of two consecutive phases. In Phase 2, the actual Active Learning component of Jasmine is executed given the results of Phase 1. More specifically, the classifier is trained on $\mathcal{L}(t)$ and applied to $\mathcal{U}(t)$ to obtain malicious probabilities (Section~\ref{subsec:trainevalpred}). Then, the informativeness measures of the unlabeled observations are calculated (Section~\ref{subsec:certanom}) and the query sample is constructed based on these measures (Section~\ref{subsec:querysample}). Next, the actual labels are provided by the human oracle. Then, the crucial part of Jasmine is performed by updating the query fractions based on the results obtained during the iteration (Section~\ref{subsec:alphadynamic}). Consequently, the query set of the next iteration should have a better balance of anomalous, uncertain and random observations. Finally, the labeled query observations are added to $\mathcal{L}(t+1)$ (Section~\ref{subsec:finalsteps}) and the procedure continues.
In Phase 1, good values of the hyperparameters of the classifier are determined (Section~\ref{subsec:tuninggbm}). Also, good values of the Jasmine-specific parameters are chosen in this phase (Section~\ref{subsec:tuningjasmine}). Since tuning of the Jasmine parameters is a reduced version of the second phase, we explain the latter first in this section.

\subsection{Classification and anomaly detection techniques}
\label{subsec:mltechniques}

\subsubsection{Gradient Boosting Machine} 
The supervised machine learning technique used as the classifier in Jasmine is the Gradient Boosting Machine (GBM), which was designed by Friedman~\cite{friedman2001greedy}. As described by Natekin et al.~\cite{natekin2013gradient}, Ogutu et al.~\cite{ogutu2011comparison}, and Carauna et al.~\cite{caruana2008empirical}, one of the main merits of the GBM is its flexibility and robustness with respect to the number of hyperparameters. It can easily be customized for different practical purposes. Another advantage of the GBM is its interpretability. Since the technique is an ensemble of multiple simple models, it can be understood more easily than a black box model. This is of great value to practitioners (in our case cybersecurity analysts). The authors do mention that one of the drawbacks of the GBM is its memory-consumption for large datasets. In cybersecurity, the datasets are rather big, so it may seem counter-intuitive to use such a tree-based ensemble method. However, the datasets on which the classifier is actually trained are in fact relatively small because of the Active Learning framework. 

\subsubsection{Isolation Forest} 
The anomaly detection method used in Jasmine is the Isolation Forest (IF), which was developed by Liu et al.~\cite{liu2008isolation}. Just as the GBM, it is a tree-based ensemble machine learning technique. Most anomaly detection models try to construct a representation of normal behavior, but the IF aims to isolate the anomalous observations. The authors use two important properties of anomalies: \textit{(i)} there are few of them, and \textit{(ii)} they have distinctly different characteristics than the majority of observations. Because these properties hold, it is relatively easy to isolate them from the rest. 
The main advantage of the IF technique is that it is fast and relatively easy to interpret~\cite{liu2012isolation}. 


\subsection{Tuning GBM hyperparameters}
\label{subsec:tuninggbm}

The GBM has hyperparameters that are determined beforehand. In Jasmine, good values are found via hyperparameter tuning. This is done on the set $\mathcal{L}(0)$, because by definition the observations in this set are the only ones with labels from the start. For each iteration of tuning, the set of hyperparameters is assigned a certain value. Then, the machine learning technique is trained on $\mathcal{L}(0)$ given this set and returns some performance metric~\cite{claesen2015hyperparameter}. In Jasmine, $k$-fold cross validation is used for training, such that each observation in $\mathcal{L}(0)$ is both utilized for training and evaluating the model. This is desirable, because $\mathcal{L}(0)$ is usually rather small, so we want to effectively use each provided label. During tuning, also the computation time is taken into account, because the GBM is retrained each AL iteration and relatively large training times quickly stack. We want to find hyperparameters that yield good performance in both predictive ability and computation time. To this end, assume there are $H$ hyperparameter combinations yielding a sequence of performance metrics $h_1, \dots, h_H$ and a sequence of computation times $t_1, \dots, t_H$. Let $h_* := \max\{h_1, \dots, h_H\}$ be the best performance and let $j_* := \argmax\{h_1, \dots, h_H\}$ be the combination yielding this performance. If there are several combinations resulting in the best performance metric, then $j_*$ is the one with the smallest computation time $t_{j_*}$. Also, if there is a combination yielding a performance of almost $h_*$, but with a much smaller computation time than $t_{j_*}$, then we prefer to go for that combination. In that case, combination $j$ is chosen as the optimal one whenever
\begin{equation*}
d_j := \frac{h_{j^*} - h_j}{t_{j^*} - t_j} < \varepsilon,
\end{equation*} 
where $\varepsilon > 0$ is some predefined threshold. If several $j$ satisfy this requirement, then the one with the smallest $d_j$ is chosen.

\subsection{Training, evaluating and predicting}
\label{subsec:trainevalpred}

The classification is done by means of a GBM, which we described in Section~\ref{subsec:mltechniques}. The hyperparameters for this technique are determined by the method that we described in Section~\ref{subsec:tuninggbm}. In iteration $t$, the GBM is trained with $k$-fold cross validation on $\mathcal{L}(t-1)$, yielding the classifier $f_t$ and some threshold probability $\theta \in (0,1)$. This $\theta$ represents the border between predicting an instance as 0 (benign) or as 1 (malicious), and is the value that maximizes some performance metric on $\mathcal{L}(t-1)$. In Jasmine, we are interested in how confident $f_t$ is in its predictions. The closer the predicted probability is to $\theta$, the less certain the model is. Intuitively, a value of $0.5$ can be seen as the least certain probability for a binary classifier, because it is precisely between 0 and 1. Therefore, $\theta$ is transformed to be $0.5$ and the probabilities are changed accordingly. We provide more reasoning why this is done in Section~\ref{subsec:certanom}. The function $\varphi_{\theta}: [0,1] \rightarrow [0,1]$, defined as
\begin{equation*}
\varphi_{\theta}(y) = \frac{(1-\theta)y}{(1-2\theta)y + \theta},
\end{equation*} 
is applied to all predicted probabilities to transform them. We chose this function, because it has the following desirable properties: \textit{(i)} $\varphi$ is continuously differentiable, \textit{(ii)} $\varphi_{\theta}(0) = 0$, \textit{(iii)} $\varphi_{\theta}(\theta) = 0.5$, \textit{(iv)} $\varphi_{\theta}(1) = 1$, and \textit{(v)} $\varphi'_{\theta}(y) \geq 0$. Note that the probabilities do not change when $\theta$ is already $0.5$: $\varphi_{\theta=0.5}(y) = y$ for all $y \in [0,1]$.


After this, $f_t$ is applied to the unlabeled set $\mathcal{U}(t-1)$ to obtain predicted probabilities $\hat{y}_u(t) \in [0,1]$ (which have been transformed by $\varphi_{\theta}$) for each $u \in \mathcal{U}(t-1)$. Consequently, the predicted class $\hat{c}_u(t)$ is $0$ if $\hat{y}_u(t) < 0.5$ and $1$ if $\hat{y}_u(t) \geq 0.5$.

\subsection{Calculating certainty score and anomaly score}
\label{subsec:certanom}

In the next step, the measures needed for the query function $\psi^{\text{Jas}}$ to determine which unlabeled observations to present to the expert are calculated. These measures are the certainty score and anomaly score.

\subsubsection{Certainty score}

The certainty score $z_u(t) \in [0,1]$ is defined as
\begin{equation}
\label{eq:uncertscore}
z_u(t) := 2\left|\hat{y}_u(t) - \frac{1}{2}\right|,
\end{equation}
for each $u \in \mathcal{U}(t-1)$. The lower this score, the more uncertain the trained model $f_t$ is about the predicted label of $u$. Equation~\eqref{eq:uncertscore} is the commonly used definition for the certainty score in AL methods, such as in ALADIN. 
It also shows why we transformed the raw predicted probabilities by $\varphi_{\theta}$. Now, the distance from $\frac{1}{2}$ can be at most $0.5$ in both the benign direction (corresponding to $\hat{y}_u(t) \downarrow 0$) and the malicious direction ($\hat{y}_u(t) \uparrow 1$), making $z_u(t)$ a symmetric score. 

\subsubsection{Anomaly score}

The IF technique that we described in Section~\ref{subsec:mltechniques} is used to determine the anomaly score $a_u(t)$ for each $u \in \mathcal{U}(t-1)$. Firstly, the class-specific observation sets are defined for each $c \in \{0, 1\}$ by 
\begin{align}
\mathcal{L}^{(c)}(t-1) &:= \{l \in \mathcal{L}(t-1): y_l = c\}, \nonumber \\
\mathcal{U}^{(c)}(t) &:= \{u \in \mathcal{U}(t-1): \hat{c}_u(t) = c\}.
\label{eq:anomsets}
\end{align}
Now, one IF is trained on the set $\mathcal{L}^{(0)}(t-1) \cup \mathcal{U}^{(0)}(t)$ and one on the set $\mathcal{L}^{(1)}(t-1) \cup \mathcal{U}^{(1)}(t)$, yielding a benign IF $I^{(0)}_t: \mathbb{R}^K \rightarrow [0,1]$ and a malicious IF $I^{(1)}_t: \mathbb{R}^K \rightarrow [0,1]$, respectively. Then, the anomaly score of $u \in \mathcal{U}(t-1)$ is defined as
\begin{equation}
\label{eq:anomscore}
a_u(t) := I^{(\hat{c}_u(t))}_t(\mathbf{x}_u).
\end{equation}
This is the output value that the appropriate IF produces when the feature vector $\mathbf{x}_u$ is fed into it.

\subsection{Constructing query sample $\mathcal{Q}(t)$}
\label{subsec:querysample}

An important component of Jasmine is the way in which the query function $\psi^{\text{Jas}}$ determines how the query sample $\mathcal{Q}(t)$ is constructed. There are three types of query observations: anomalies, uncertainties and random instances. Which part of $\mathcal{Q}(t)$ should be allocated to which type is given by the anomaly fraction $\alpha_a(t)$, uncertainty fraction $\alpha_z(t)$ and randomness fraction $\alpha_r(t)$. 

\subsubsection{Anomalous query observations}
For each class $c \in \{0, 1\}$, the unlabeled observations in $\mathcal{U}^{(c)}(t)$ (see~\eqref{eq:anomsets}) are sorted from most anomalous to least anomalous. Then, the top $\frac{1}{2}\alpha_a(t)\cdot Q$ observations are taken as the anomalous query instances for predicted class $c$.

\subsubsection{Uncertain query observations}
The uncertainties are selected in a similar way: the observations in $\mathcal{U}^{(c)}(t)$ are sorted from least certain to most certain for each class $c$. Then, the top $\frac{1}{2}\alpha_z(t) \cdot Q$ instances are selected for the uncertain query observations for predicted class $c$.

\subsubsection{Random query observations}
The random query observations are selected in a simple way: a sample of size $\alpha_r(t) \cdot Q$ is taken from $\mathcal{U}(t)$ (without the already selected anomalous and uncertain observations).

\subsubsection{Corrections}
First of all, some rounding corrections are applied when the values $\frac{1}{2}\alpha_a(t)\cdot Q$, $\frac{1}{2}\alpha_z(t)\cdot Q$ and/or $\alpha_r(t)\cdot Q$ are not integer. Secondly, it is possible that there are not enough observations for a specific class. Then, observations from the other class are added to reach the number of required anomalies or uncertainties. Thirdly, there can be an overlap between the most anomalous and least certain instances: some observations can be both anomalous and uncertain, and hence, we select them for both query types.

Finally, when the query observations are determined, the query set $\mathcal{Q}(t)$ is shown to the human expert and they provide the correct label $y_q$ for each $q \in \mathcal{Q}(t)$.

\subsection{$\alpha$-dynamic update}
\label{subsec:alphadynamic}

\subsubsection{Constructing update parameters}
The query set $\mathcal{Q}(t)$ obtained in Section~\ref{subsec:querysample} can be decomposed in $\mathcal{Q}_a(t)$, $\mathcal{Q}_z(t)$ and $\mathcal{Q}_r(t)$, which are the sets of anomalous, uncertain and random queries, respectively. Also, let $Q_a(t)$, $Q_z(t)$ and $Q_r(t)$ be their respective sizes. 
Furthermore, let $y_q \in \{0,1\}$ be the real label of query observation $q$, $\hat{y}_q(t) \in [0,1]$ its predicted malicious probability and $\hat{c}_q(t) \in \{0,1\}$ its predicted label. In practice, the real label is available since the cyber expert provides it. We want to construct a metric for the anomalous queries and one for the uncertain queries that describe how much information the observations of those types can potentially add to the model. We do this by looking at the false negatives (FNs) and false positives (FPs) in $\mathcal{Q}(t)$. The reasoning is that when there are many FNs and FPs, then the model was bad at predicting the real classes of those unlabeled observations. Hence, adding these observations to the labeled set $\mathcal{L}(t)$ could yield a lot of information for the classifier trained in the next iteration. In short, we consider the fraction of FPs and FNs in $\mathcal{Q}(t)$, but also take a look at how convincing they are. If the model is fairly certain that an observation is malicious while it is in fact benign, then this FP obtains a larger weight than if the model is not that sure. Consequently, we define the anomaly information metric $\delta^{\beta}_a(t)$ as follows:
\begin{equation}
\label{eq:anomdelta}
\delta^{\beta}_a(t) := \frac{\sum_{q \in \mathcal{Q}_a(t)} |\hat{y}_q(t) - y_q| \cdot \left(\beta \cdot \mathbf{1}_{\{\hat{c}_q = 0, y_q = 1\}} + \mathbf{1}_{\{\hat{c}_q = 1, y_q = 0\}}\right) }{Q_a(t) + (\beta - 1)\cdot|\{q \in \mathcal{Q}_a(t): y_q = 1\}|}.
\end{equation}
Note that the first indicator function in~\eqref{eq:anomdelta} corresponds to an FN: the real label is 1, but the predicted label is 0. Equivalently, the second indicator function corresponds to an FP. If the query observation $q$ is an FN, it gets weighted by some parameter $\beta > 0$. This enables us to put more ($\beta > 1$) or less ($\beta < 1$) emphasis on the FNs compared to the FPs. Thus, if $q$ is an FN, then it obtains value $\beta|\hat{y}_q(t) - y_q|$; if it is an FP, it obtains $|\hat{y}_q(t) - y_q|$. The farther the predicted probability $\hat{y}_q(t)$ is from 0 or 1 (thus the less certain the model is about the prediction), the larger the assigned value for $q$ becomes. The denominator ensures that the value of $\delta^{\beta}_a(t)$ is at most 1. When there are no FPs and FNs, then the value of the metric is 0. Consequently, $\delta^{\beta}_a(t) \in [0,1]$. The larger $\delta^{\beta}_a(t)$ is, the more information the anomalies convey, because we expect that incorrectly predicted observations add relevant information to the model if they are added to the train set with the correct labels. 

Similarly, we define the uncertainty information metric
\begin{equation}
\label{eq:uncertdelta}
\delta^{\beta}_z(t) := \frac{\sum_{q \in \mathcal{Q}_z(t)} |\hat{y}_q(t) - y_q|\cdot\left(\beta \cdot \mathbf{1}_{\{\hat{c}_q = 0, y_q = 1\}} + \mathbf{1}_{\{\hat{c}_q = 1, y_q = 0\}}\right)}{Q_z(t) + (\beta - 1)\cdot|\{q \in \mathcal{Q}_z(t): y_q = 1\}|},
\end{equation}
as a measure on how much information the uncertain observations in the query set add on average. Also, $\delta^{\beta}_z(t) \in [0,1]$. 

Next, the difference between the information metrics that we defined in~\eqref{eq:anomdelta} and~\eqref{eq:uncertdelta} describes whether anomalies or uncertainties could add more information to the model. We define this difference $\Delta(t)$ as
\begin{equation*}
\Delta(t) := \delta^{\beta}_a(t) - \delta^{\beta}_z(t) \in [-1, 1].
\end{equation*}
It is used to determine how the query fractions of anomalies and uncertainties are updated. If $\Delta(t) > 0$, then the queried anomalies could add more information on average, and hence, preferably, more anomalies are selected the next iteration. If $\Delta(t) < 0$, then the uncertainties could add more information, and so, more uncertainties are selected. 

Now, we want the possibility to put more or less emphasis on bigger or smaller values of $\Delta(t)$. Hence, we introduce a non-linearity governed by $\gamma > 0$ to obtain $\Delta^{\gamma}(t)$. To this end, we define the update factor as  
\begin{equation}
\label{eq:deltagamma}
\Delta^{\gamma}(t) := \text{sgn}(\Delta(t)) \cdot |\Delta(t)|^{1/\gamma}.
\end{equation}
When $\gamma = 1$, then~\eqref{eq:deltagamma} reduces to the linear case $\Delta^{\gamma}(t) = \Delta(t)$. If $0 < \gamma < 1$, then $|\Delta^{\gamma}(t)| \leq |\Delta(t)|$, so the update factor is relatively smaller. On the other hand, if $\gamma > 1$, then $|\Delta^{\gamma}(t)| \geq |\Delta(t)|$ and the factor is relatively larger.

\subsubsection{Defining $\alpha_a(t+1)$ and $\alpha_z(t+1)$}
The update factor $\Delta^{\gamma}(t)$ is used to determine whether more anomalies or uncertainties should be queried in the next iteration such that we obtain a better predictive model. The updates $\alpha_a(t+1)$ and $\alpha_z(t+1)$ have the forms
\begin{equation}
\label{eq:prealphaanom}
\alpha_a(t+1) =  {\lambda}_{t+1}\left(\alpha_a(t) + w_a^{(1)}(t) \cdot \max\{0, \Delta^{\gamma}(t)\} + w_a^{(2)}(t) \cdot \min\{0, \Delta^{\gamma}(t)\}\right),
\end{equation}
and
\begin{equation}
\label{eq:prealphauncert}
\alpha_z(t+1) =  {\lambda}_{t+1}\left(\alpha_z(t) + w_z^{(1)}(t) \cdot \max\{0, -\Delta^{\gamma}(t)\} + w_z^{(2)}(t) \cdot \min\{0, -\Delta^{\gamma}(t)\}\right),
\end{equation}
respectively. Here, the $w$ variables are non-negative constants that we need to ensure that the fractions stay within the correct bounds for time step $t$. Moreover, $\lambda_{t+1}$ denotes a linear function which guarantees that the updated fractions are also within the bounds for time step $t+1$. Let us examine~\eqref{eq:prealphaanom} in a bit more detail: the new fraction $\alpha_a(t+1)$ is based on the old value $\alpha_a(t)$ plus some value when $\Delta^{\gamma}(t) > 0$ or minus some value when $\Delta^{\gamma}(t) < 0$. In other words, when the anomalies add more information on average than the uncertainties, then $\alpha_a(t+1)$ becomes larger. Otherwise, $\alpha_a(t+1)$ becomes smaller. The update dynamics are the other way around for~\eqref{eq:prealphauncert}. We provide the explicit mathematical definitions of the $w$ variables and $\lambda_{t+1}$ in~\ref{sec:appendix}. It is important to mention that they depend on the hyperparameters $\alpha_a^{(0)}$ and $\alpha_z^{(0)}$, which are the initial query fractions of anomalies and uncertainties, respectively.

\subsubsection{Defining $\alpha_r(t+1)$}
The fraction of random observations that we want to query relates to the number of available labeled observations $L(t)$. If $L(0)$ is small, then $\mathcal{L}(0)$ is less likely to be a good representation of the complete labeled dataset $(\mathbf{X}, \mathbf{y})$. In that case, we want to query relatively more random instances at the start to be able to obtain a representative train set for the GBM classifier to learn from. As $L(t)$ increases, we want to query less and less random observations and let Active Learning take over in the sense of querying anomalies and uncertainties. Therefore, let $\alpha_r(t)$ be an exponentially decreasing function in $t$. More specifically,
\begin{equation}
\alpha_r(t) = \alpha^{\max}_r \cdot 2^{-\tau \cdot L(t)},
\label{eq:prealpharand}
\end{equation}
with $\tau > 0$ the decrease speed and $L(t) = L(0) + Q \cdot t$. Note that this function is determined beforehand. Hence, how the fraction of randomly queried observations changes, is fixed during the Jasmine procedure. The value of $\alpha^{\max}$ is directly determined by the hyperparameters $\alpha_a^{(0)}$ and $\alpha_z^{(0)}$.


\subsubsection{Computation time}
It is important to see that executing the $\alpha$-dynamic update step does not take much computation time. No additional machine learning models have to be trained and only relatively simple operations are performed to determine the values of the query fractions for the next time step.

\subsection{Final iteration updates}
\label{subsec:finalsteps}

The last steps that Jasmine has to perform are rather simple, the labeled query set $\mathcal{Q}(t)$ is added to the current labeled set $\mathcal{L}(t-1)$ and removed from the current unlabeled set $\mathcal{U}(t-1)$. More specifically, $\mathcal{L}(t) := \mathcal{L}(t-1) \cup \mathcal{Q}(t)$ and $\mathcal{U}(t) := \mathcal{U}(t-1) \setminus \mathcal{Q}(t)$.

\subsection{Tuning Jasmine hyperparameters}
\label{subsec:tuningjasmine}

Chronologically speaking, tuning of the Jasmine-specific hyperparameters takes place in Phase 1 before the actual Active Learning procedure in Phase 2, as we illustrated in Figure~\ref{fig:jasmine}. To be more specific, \textit{Jasmine tuning} occurs after tuning the hyperparameters for the GBM, and thus, directly after Section~\ref{subsec:tuninggbm}. 
Recall that the hyperparameters are \textit{(i)} $\alpha_a^{(0)}$ and $\alpha_z^{(0)}$, the starting fractions of querying anomalous and uncertain observations, respectively; \textit{(ii)} $\beta$, the parameter assigning a certain weight to an FN compared to an FP, given in~\eqref{eq:anomdelta} and~\eqref{eq:uncertdelta}; \textit{(iii)} $\gamma$, the update magnitude, given in~\eqref{eq:deltagamma}; and \textit{(iv)} $\tau$, the decrease speed in querying random observations, given in~\eqref{eq:prealpharand}. Using normalization, $\alpha_a^{(0)}$ and $\tau$ completely determine $\alpha_z^{(0)}$, so the latter does not have to be tuned.

To determine appropriate values for these hyperparameters, the initially labeled set $\mathcal{L}(0)$ is randomly partitioned into the sets $\mathcal{L}_J(0)$, $\mathcal{U}_J(0)$ and $\mathcal{E}_J$. During Jasmine tuning, $\mathcal{L}_J(0)$ is taken as the initially labeled set, $\mathcal{U}_J(0)$ as the initially unlabeled set and $\mathcal{E}_J$ as the evaluation set. Let $\mathcal{J}$ be the set with hyperparameter values that we want to consider for tuning. Thus, an element $j \in \mathcal{J}$ is a four-dimensional vector of the form $(\alpha_a^{(0)}, \beta, \gamma, \tau)$. 
Let $Q_J$ be the number of unlabeled observations that should be queried in each iteration. Ideally, the value of $Q_J$ is close to $Q$, but we do want to perform some iterations before $\mathcal{L}_J(t)$ reaches the size of $\mathcal{L}(0)$. The number of iterations in Jasmine tuning is given by $T_J := \lceil \frac{U_J(0)}{Q_J} \rceil - 1$, where $U_J(0) := |\mathcal{U}_J(0)|$. The minus one is because the last unlabeled observations are the least informative, and hence, not of interest to query. 

Let $j \in \mathcal{J}$ be some hyperparameter combination. Then a GBM model is trained on $\mathcal{L}_J(0)$, similar to what we described in Section~\ref{subsec:trainevalpred}. This model is then applied to the evaluation set $\mathcal{E}_J$ resulting in some performance metric $p_j(0)$. Then the rest of the AL procedure, as we described in Section~\ref{subsec:certanom} to \ref{subsec:finalsteps}, is executed. After all iterations are performed, the sequence of performance measures $\{p_j(t-1)\}_{\{t = 1, \dots, T_J\}}$ is obtained. We use this sequence to determine which hyperparameter combination works best, since such a sequence is obtained for each $j \in \mathcal{J}$.

Because stochasticity is involved, we repeat Jasmine tuning $S_J$ times. Each simulation, the set $\mathcal{L}(0)$ is randomly divided in the sets $\mathcal{L}_J(0)$, $\mathcal{U}_J(0)$ and $\mathcal{E}_J$. Also, each simulation yields a sequence of performance metrics for each hyperparameter combination. Then the combination that yields the best performance over the simulations is taken as (relatively) optimal for $\alpha_a^{(0)}$, $\beta$, $\gamma$ and $\tau$.

\subsection{Summary of Jasmine}

\begin{algorithm} 
\caption{Jasmine procedure} 
\label{alg:jasmine} 
\begin{algorithmic}[1] 
\REQUIRE Labeled set $\mathcal{L}(0)$, unlabeled set $\mathcal{U}(0)$, number of iterations $T$, query function $\psi^{\text{Jas}}$
\STATE Tune parameters of GBM on $\mathcal{L}(0)$ (Section~\ref{subsec:tuninggbm})
\STATE Tune Jasmine-specific hyperparameters $(\alpha_a^{(0)}, \beta, \gamma, \tau)$ on $\mathcal{L}(0)$ (Section~\ref{subsec:tuningjasmine}) and determine $\alpha_z^{(0)}$ by normalization
\FOR{$t = 1$ to $T$}
    \STATE Train GBM $f_t$ with tuned parameters on $\mathcal{L}(t-1)$ 
    \STATE Apply $f_t$ to $\mathcal{U}(t-1)$ to obtain predictions $\hat{y}_u(t)$
    \STATE Assign each $u \in \mathcal{U}(t-1)$ to its most likely class
    \STATE Calculate each $z_u(t)$ as defined in~\eqref{eq:uncertscore}
    \STATE Construct one IF for each class (Section~\ref{subsec:certanom}).
    \STATE Compute each $a_u(t)$ as defined in~\eqref{eq:anomscore}
    \STATE Compose $\mathcal{Q}(t)$ using query function $\psi^{\text{Jas}}$ as described in Section~\ref{subsec:querysample}
    \STATE Obtain actual classes $y_q$ of $q \in \mathcal{Q}(t)$ by human expert
    \STATE Use $y_q$ and predictions $\hat{y}_q(t)$ to determine $\delta_a^{\beta}(t)$,  $\delta_z^{\beta}(t)$ and $\Delta^{\gamma}(t)$ with~\eqref{eq:anomdelta},~\eqref{eq:uncertdelta} and~\eqref{eq:deltagamma}, respectively.
    \STATE Update query fractions to obtain $\alpha_a(t+1)$ and $\alpha_z(t+1)$ (Section~\ref{subsec:alphadynamic})
    \STATE $\mathcal{L}(t) \Leftarrow \mathcal{L}(t-1) \cup \mathcal{Q}(t)$ and $\mathcal{U}(t) \Leftarrow \mathcal{U}(t-1) \setminus \mathcal{Q}(t)$ 
\ENDFOR
\RETURN
\end{algorithmic}
\end{algorithm}

The complete Jasmine procedure is summarized in Algorithm~\ref{alg:jasmine}.

\section{Experimental Setup}
\label{sec:expsetup}

We conducted several experiments to determine whether our AL method Jasmine performed better than ALADIN and to decide if $\alpha$-dynamic updating yielded significant improvements over baseline query functions. In this section, firstly, we discuss the datasets on which the experiments were performed. These sets are fully labeled, so the labels for the observations in the `unlabeled' set were hidden until they were queried by Jasmine. After that, we explain the steps taken to execute the procedures. These include which hyperparameters were tuned over which ranges and how the different AL methods were evaluated.

\subsection{Data}

Yavanoglu et al.~\cite{yavanoglu2017review} and Ferrag et al.~\cite{ferrag2020deep} provided overviews of publicly available security-related datasets commonly used in intrusion detection research. The sets that were discussed span from 1999 to 2018, showing that even old network datasets are still used to benchmark intrusion detection techniques. However, this is mostly due to the lack of public data, as discussed in Section~\ref{sec:intro}. Here, firstly, we used the \textit{NSL-KDD} dataset for assessing the considered AL methods, since it is the most used for evaluation in the field of cybersecurity. Ferrag et al.\ state that this data was cited at least $1{,}630$ times as of June 22, 2019. Also, Stokes et al.\ made use of the NSL-KDD dataset to assess their ALADIN method. Secondly, we considered the \textit{UNSW-NB15} dataset. This set is cited often too, namely $202$ times, and it is much more recent than the popular NSL-KDD data. Moreover, it has several more realistic aspects, which are discussed later. Henceforth, these datasets were used to assess the performance of the discussed AL methods and to compare the obtained results for different data.

\subsubsection{NSL-KDD}
The NSL-KDD dataset was developed by Tavallaee et al.\ in 2009~\cite{tavallaee2009detailed} and is an improvement on the KDD-Cup-99 dataset. The latter was prepared by Stolfo et al.~\cite{stolfo2000cost} and consists of a train set and a test set. These two sets were constructed such that they do not have the same underlying distribution. 
Each observation in KDD-Cup-99 is made up of a 41-dimensional feature vector and an output label with the attack type. There are four global types of actual attacks and a `normal' type, indicating a benign connection. Within the attack types, there can be several distinct attack scenarios. The test set contains scenarios from the train set as well as new scenarios.
As mentioned before, the NSL-KDD dataset is a revision of the KDD-Cup-99 data. Firstly, redundant and invalid records were removed to ensure learning algorithms do not get biased towards frequent records. 
Secondly, specific records were selected from the resulting set to construct a more challenging dataset. How this was done is described by Tavallaee et al.~\cite{tavallaee2009detailed}. Finally, the total number of train observations in NSL-KDD is $125{,}972$ and the number of test instances is $22{,}544$. We removed the categorical features \textsf{Protocol\_type}, \textsf{Service}, \textsf{Flag}, \textsf{Difficulty\_level}, because Jasmine is based on numerical techniques. Moreover, since Jasmine expects binary output labels, all attack types obtained value 1 (the malicious class), while the normal type obtained value 0 (the benign class). $46.5\%$ of the train instances are malicious, while $56.9\%$ are malicious in the test set.

\subsubsection{NSL-KDD-rand}
Besides considering the NSL-KDD data as provided by Tavallaee et al., we also considered the dataset we call \textit{NSL-KDD-rand}. We constructed this set by combining the train and test set of the NSL-KDD dataset. Now, $48.1\%$ of all observations are malicious. During the experiments, each time a new train and test set were chosen. These new sets are expected to have the same underlying distribution, in contrast to the provided train and test set of the NSL-KDD data. It is interesting to see how the considered AL methods performed on differently structured data.

\subsubsection{UNSW-NB15}
The UNSW-NB15 dataset was constructed by Moustafa et al.\ in 2015~\cite{moustafa2015unsw}, partially to address some problems of the NSL-KDD data. The authors used the so-called IXIA PerfectStorm tool to generate a mix of realistic modern benign network behaviors and synthetic malicious connections. The dataset contains $2{,}540{,}047$ observations with each connection consisting of a $47$-dimensional feature vector and two output attributes. The first output is the specific attack type and the second is a binary value indicating whether the observation is benign (0) or malicious (1). Nine attack types are present in the dataset, but we only used the second output attribute, since Jasmine expects a binary output label. We reduced the number of predictive features from $47$ to $36$ by removing \textsf{srcip}, \textsf{sport}, \textsf{dstip}, \textsf{dsport}, \textsf{proto}, \textsf{state}, \textsf{service}, \textsf{stcpb}, \textsf{dtcpb}, \textsf{Stime} and \textsf{Ltime}, because they directly determine the output label, are categorical or have no predictive use. Furthermore, there are several missing values in UNSW-NB15 which we chose from context to be 0. Finally, $12.6\%$ of the observations correspond to attacks. This lower fraction of malicious traffic is one of the reasons why this dataset is closer to reality than the NSL-KDD data. Another reason is that this data does contain modern attack styles and modern benign traffic behavior. More information about UNSW-NB15 is given by Moustafa et al.~\cite{moustafa2015unsw}.

\subsection{Experiments}

\subsubsection{Query functions} 
We considered several query functions in the experiments. First of all, we regarded Jasmine with its characteristic $\alpha$-dynamic query function $\psi^{\text{Jas}}$ (as described in Section~\ref{subsec:querysample}) as the main focus of this research ({\textproc{jas.main}}). Furthermore, we also examined some simpler query functions for the Jasmine procedure: only querying anomalies ({\textproc{jas.anom}}), only querying uncertainties ({\textproc{jas.uncert}}), and only querying random observations ({\textproc{jas.rand}}). Note that for these three query functions $\alpha$-dynamic updating is not involved, meaning that the corresponding procedures are incomplete versions of Jasmine. 
It is interesting to see what the results were for only querying one specific type of query observations compared to querying a mix of them. Naturally, we also considered the full ALADIN procedure of Stokes et al.\ ({\textproc{ala.main}}), just as the incomplete Jasmine procedure with ALADIN's query function of querying anomalies and uncertainties in a fixed 50/50 split ({\textproc{jas.basic}}). Consequently, we executed six different AL methods.

\subsubsection{Global parameters}
The global parameters are the variables defined before any computation took place. These include the initial size of the labeled set $L(0)$, the initial size of the unlabeled set $U(0)$, the size of the evaluation set $E$ and the query set size $Q$. Moreover, $N$ is the maximum number of observations that were to be queried to the human expert during the process. Since the goal of AL is to label as few observations as possible, we were not interested in querying every unlabeled observation. Together with $Q$, the parameter $N$ determines the total number of iterations: $T := \lfloor N/Q \rfloor$. Finally, since Jasmine is an inherently stochastic procedure, we repeated the experiments $S$ times. This was done to analyze how our method behaved on average and in what range its performances resided. Note that, for each repetition, the sets $\mathcal{L}(0)$ and $\mathcal{U}(0)$ were newly constructed. For the NSL-KDD-rand and UNSW-NB15 datasets, the evaluation set $\mathcal{E}$ was also freshly sampled. This was not necessary for the NSL-KDD data, since $\mathcal{E}$ is a provided fixed set. 

The values chosen for the global parameters for the experiments are presented in Table~\ref{tab:globalparam1} and~\ref{tab:globalparam2}. As the tables show, two different initial labeled data set sizes were considered, because we were interested in how $L(0)$ influenced the performance of the AL methods. Furthermore, we chose $Q$ to be 40, as we deemed this a good balance between allowing for the query fractions to update not too erratically (needing $Q$ to be large) and allowing for relatively small updates to the classifier (needing $Q$ to be small). Also, we chose $N$ to be $15{,}000$ because we saw from exploratory studies that the models do not drastically change anymore when more labels were provided. Finally, for the NSL-KDD-rand and UNSW-NB15 datasets, we chose $E$ to be $5{,}000$, to obtain a representative test set without using too many observations only for evaluation.
\begin{table}[h!]
  \begin{center}
    \caption{Values for global parameters on NSL-KDD}
    \label{tab:globalparam1}
    \begin{tabular}{c|c|c|c|c|c} 
      $L(0)$ & $U(0)$ & $E$ & $Q$ & $N$ & $S$ \\
      \hline
      $125$ & $125{,}848$ & $22{,}544$ & $40$ & $15{,}000$ & $30$ \\
      $250$ & $125{,}723$ & $22{,}544$ & $40$ & $15{,}000$ & $30$
   \end{tabular}
  \end{center}
\end{table}
\begin{table}[h!]
\begin{center}
    \caption{Values for global parameters on NSL-KDD-rand and UNSW-NB15}
    \label{tab:globalparam2}
    \begin{tabular}{c|c|c|c|c|c} 
      $L(0)$ & $U(0)$ & $E$ & $Q$ & $N$ & $S$ \\
      \hline
      $125$ & $146{,}392$ & $5{,}000$ & $40$ & $15{,}000$ & $30$ \\
      $250$ & $146{,}267$ & $5{,}000$ & $40$ & $15{,}000$ & $30$
    \end{tabular}
  \end{center}
\end{table}

\subsubsection{Hyperparameter tuning GBM} 
As mentioned in Section~\ref{subsec:tuninggbm}, good values for the GBM were found by tuning on the initially labeled set $\mathcal{L}(0)$ with $k$-fold cross validation. We used the study by Tama et al.~\cite{tama2019depth} and exploratory research to determine which hyperparameters to tune and over what range to tune them. The parameters that were not selected for tuning obtained their default settings as given by the \texttt{h2o.gbm} function of the \texttt{H2O.ai} package, which we used in the \texttt{R} programming language. The values or tuning ranges of the parameters are shown in Table~\ref{tab:gbmparam}. Since the number of possible hyperparameter combinations is large (more than $177$ thousand), tuning was performed using a random search over all combinations for a maximum of 4 hours. The combination with the best trade-off between performance metric and computation time was chosen. 
We considered the $F_1$ score as the performance metric and we chose the threshold $\varepsilon$ to be $10^{-4}$.
\begin{table}[h!]
  \begin{center}
    \caption{Tuning ranges for hyperparameters in \texttt{h2o.gbm} (\textsf{csr} = `col\_sample\_rate')}
    \label{tab:gbmparam}
    \begin{tabular}{c|c|c} 
      distribution & histogram\_type & learn\_rate\_annealing   \\
      Bernoulli & RoundRobin &  $\{0.95, 0.99, 0.999\}$ \\
      \hline
      max\_depth & sample\_rate & ntrees   \\
     $\{6, 12, 24\}$ & $\{0.60, 0.78, 1.0\}$ & $\{250, 500, 1{,}000\}$  \\
      \hline
    nbins & nbins\_cats & learn\_rate  \\
     $\{10, 16, 25\}$ & $\{16, 32, 64\}$ & $\{0.02, 0.05, 0.125\}$  \\
      \hline
     min\_rows & \textsf{csr} & \textsf{csr}\_change\_per\_level   \\
     $\{6, 8, 10\}$ & $\{0.84, 0.92, 1.0\}$ & $\{0.94, 1.0, 1.06\}$ \\
     \hline
     &\textsf{csr}\_per\_tree & \\
     & $\{0.40, 0.64, 1.0\}$ & 
        \end{tabular}
  \end{center}
\end{table}
\begin{table}[h!]
\begin{center}
    \caption{Tuning ranges for Jasmine parameters}
    \label{tab:jasparam}
    \begin{tabular}{c|c|c|c|c} 
      $\alpha_a^{(0)}$ & $\beta$ & $\gamma$ & $\tau$ & $S_J$ \\
      \hline
      $\left\{\frac{1}{4}, \frac{1}{2}, \frac{3}{4}\right\}$ & $\left\{\frac{1}{2}, 1, 2\right\}$ & $\left\{\frac{1}{2}, 1, 2\right\}$ & $\left\{\frac{1}{800}, \frac{1}{400}, \frac{1}{200}, \frac{1}{100}\right\}$ & 4\\
    \end{tabular}
  \end{center}
\end{table}

\subsubsection{Jasmine tuning} 
The relevant hyperparameters for the Jasmine tuning phase, as we described in Section~\ref{subsec:tuningjasmine}, are the initial anomaly query fraction $\alpha_a^{(0)}$, the FN weight factor $\beta$, the update magnitude $\gamma$ and the decrease speed in querying random instances $\tau$. The ranges that the parameters were tuned over are shown in Table~\ref{tab:jasparam}, yielding 108 possible combinations. We chose the ranges for $\alpha_a^{(0)}$, $\beta$ and $\gamma$ to be symmetric around the `unity value'. For $\alpha_a^{(0)}$, this is $\frac{1}{2}$, since then the initial number of anomalous observations was equal to the number of uncertain instances. For $\beta$, this is $1$, because then the FNs and FPs were weighed equally. For $\gamma$, this is also $1$, since then $\Delta^{(\gamma)}(t)$ reduced to the linear difference $\Delta(t)$.

During Jasmine tuning, we wanted to choose $Q_J$ as close to $Q$ as possible, but also perform at least three iterations. Hence, we defined the tuning query size as $Q_J := \min\{U_J(0)/4, Q\}$. The initially unlabeled set size was divided by 4 ($=3+1$), since the last iteration was not performed. This is because we deem the last unlabeled observations the least informative. Each Jasmine parameter combination $j$ yielded a performance metric for every time step. This produced the sequence $\{p_j(t)\}_{\{t = 0, \dots, T_J - 1\}}$. Again, this metric was the $F_1$ score. After the iterations were performed, the area underneath the $(t, p_j(t))_{\{t = 0, \dots, T_J - 1\}}$-`curve' was calculated. This is an example of a \textit{learning curve}, which is commonly used in the AL paradigm to assess the quality of a method~\cite{settles2009active, kumar2020active}. The larger this area, the better parameter combination $j$ is. As there is stochasticity in the techniques used in Jasmine, the tuning phase was repeated $S_J$ times. The combination with the largest average area was chosen as the hyperparameter setting for Jasmine in the actual AL procedure. Note that the Jasmine-specific parameters were tuned on $\mathcal{L}(0)$, and so, Jasmine did not get an unfair advantage by seeing more data in advance than the other AL methods, which did not need to execute Jasmine tuning.

\subsubsection{Evaluation of AL methods} 
To evaluate the six different AL methods, we utilized the evaluation set $\mathcal{E}$ that was set aside each simulation. The quality of the predictions of an AL method on $\mathcal{E}$ was determined by the performance metric $p(t)$ (in this case the $F_1$ score) for every iteration $t$. After some reference value $t_{\text{ref}}$ of iterations were performed ($0 \leq t_{\text{ref}} \leq T$), a sequence of performance metrics $\{p(t)\}_{\{t = 0, \dots, t_{\text{ref}}\}}$ was obtained. Similar to Jasmine tuning, we took the area $A(t_{\text{ref}})$ underneath the $(t, p(t))_{\{t = 0, \dots, t_{\text{ref}}\}}$-learning curve as a measure of performance. Briefly said, the higher $A(t_{\text{ref}})$, the better the method is up to the iteration step $t_{\text{ref}}$. However, since there is stochasticity involved, we repeated each complete AL procedure $S$ times. During simulation $s$, $\mathcal{L}(0)$ and $\mathcal{U}(0)$ were randomly chosen (for NSL-KDD-rand and UNSW-NB15 also $\mathcal{E}$ was randomly sampled) and all six procedures were provided the same initial sets. Consequently, this led to the vector $(A^{(1)}_s(t_{\text{ref}}), \dots, A^{(6)}_s(t_{\text{ref}}))$ of paired area metrics. 
Next, we statistically compared the area metrics of the Jasmine $\alpha$-dynamic method with the metrics of the other methods. This was done by the Wilcoxon signed-rank test (with significance threshold of $0.05$) to determine whether 
\begin{equation}
H_0^{(m)}(t_{\text{ref}}): \underset{s = 1, \dots, S}{\text{median}} \left\{A^{(1)}_s(t_{\text{ref}})\right\} < \underset{s = 1, \dots, S}{\text{median}} \left\{A^{(m)}_s(t_{\text{ref}})\right\}
\label{eq:wilcox}
\end{equation}
could be rejected for method $m = 2, \dots, 6$. When this was the case, then $\alpha$-dynamic querying performed significantly better than the other considered query functions and AL techniques.

\section{Results}
\label{sec:results}

In this section, the results of our research are presented. They were obtained by performing the steps explained in Section~\ref{sec:expsetup} on the NSL-KDD, NSL-KDD-rand and UNSW-NB15 data. First of all, the learning curve of $F_1$ scores is shown for each of the six AL methods that we considered. This curve gives an insight in how well the classifier of a specific method performed on the evaluation set throughout the AL process. Secondly, the $p$-values of the Wilcoxon signed-rank test are presented for predetermined specific iteration steps. 
Thirdly, the dynamics of the query fractions $\alpha_a(\cdot)$, $\alpha_z(\cdot)$ and $\alpha_r(\cdot)$ are shown to illustrate how Jasmine adjusted the balance between querying anomalous, uncertain and random observations. Finally, we discuss the implications of these results per dataset.

\subsection{Results on NSL-KDD}
\label{subsec:nslkdd}

\subsubsection{Learning curves}
\begin{figure}[h!]
\centering
\begin{subfigure}[b]{.65\textwidth}
 \centering
 \includegraphics[width=1\linewidth]{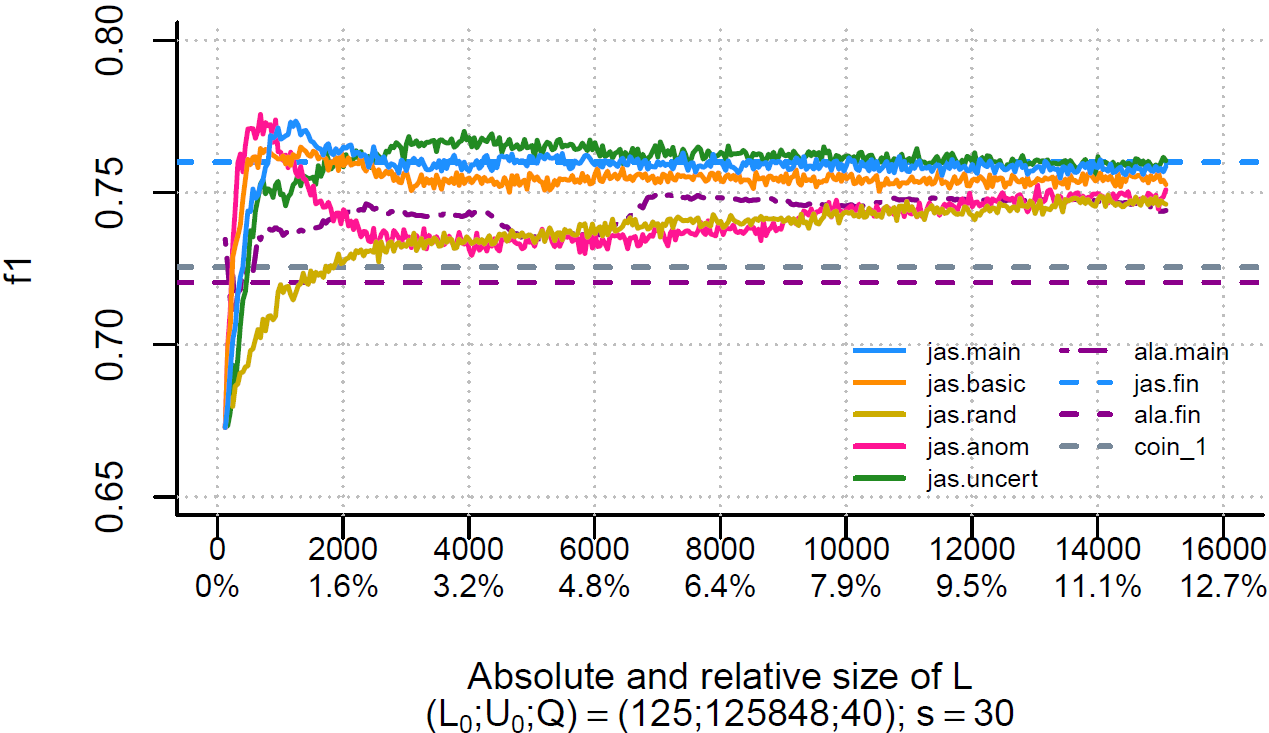}
 \caption{$L(0) = 125$}
 \label{fig:f1l125nslkdd}
\end{subfigure}

\begin{subfigure}[b]{.65\textwidth}
 \centering
 \includegraphics[width=1\linewidth]{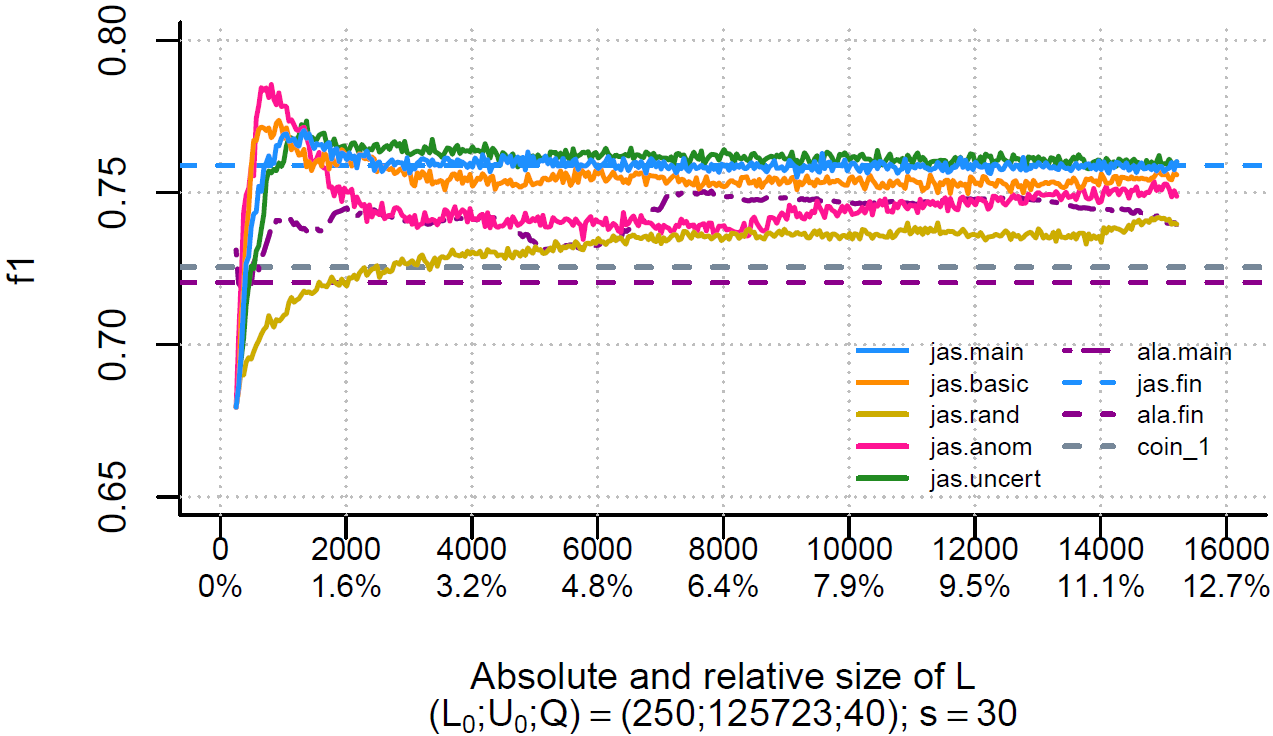}
 \caption{$L(0) = 250$}
 \label{fig:f1l250nslkdd}
\end{subfigure}
\caption{Learning curves on NSL-KDD-$\mathcal{E}$ for different initial sizes $L(0)$.}
\label{fig:f1nslkdd}
\end{figure}

Figure~\ref{fig:f1nslkdd} shows the average learning curves for each of the six AL methods on the fixed evaluation set NSL-KDD-$\mathcal{E}$ for initially labeled set sizes $L(0) = 125$ and $L(0) = 250$. Each simulation yielded a learning curve, hence we took the average of the curves over all simulations. 
The blue dashed line ({\textproc{jas.fin}}) is the average final performance on $\mathcal{E}$ of the GBM trained on the complete train set with all labels available. This performance metric was $\overline{F_1} \approx 0.760$ for Figure~\ref{fig:f1l125nslkdd} and $\overline{F_1} \approx 0.759$ for Figure~\ref{fig:f1l250nslkdd}. 
The value of the line is approximately the same for both plots, because the evaluation set $\mathcal{E}$ is fixed and independent of $\mathcal{L}(0)$. However, each GBM was tuned differently, leading to small differences. 
Since all labels of the NSL-KDD dataset are technically available, we included the final performance to show how quickly the Jasmine-based AL methods reached this value. The purple dashed line ({\textproc{ala.fin}}) is the final performance on $\mathcal{E}$ of the logistic regression classifier of the ALADIN procedure ($F_1 \approx 0.720$). This final performance was constant during the simulations and for both settings of $L(0)$, since the classifier of ALADIN is deterministic. Finally, the gray dashed line ({\textproc{coin\_1}}) is the expected performance on $\mathcal{E}$ ($F_1 \approx 0.725$) of the best dummy classifier, which classifies each evaluation observation as malicious. 

\subsubsection{Statistical tests}
\begin{table}[h!]
  \begin{center}
    \caption{$p$-values Wilcoxon test {\textproc{jas.main}} vs.\ \dots with $L(0) = 125$}
    \label{tab:pl125nslkdd}
    \begin{tabular}{c|c|c|c|c|c|c} 
      $t_{\text{ref}}$ & $L(t_{\text{ref}})$ & {\textproc{jas.basic}} & {\textproc{jas.rand}} & {\textproc{jas.anom}} & {\textproc{jas.uncert}} & {\textproc{ala.main}} \\
      \hline
      $9$ & $485$ & \sig{0.992} & \sig{0.000172} & \sig{1.00} & \sig{0.0155} & \sig{0.894} \\
      $16$ & $765$ & \sig{0.991} & {\color{forestgreen}$1.52\cdot10^{-5}$} & \sig{1.00} & \sig{0.0249} & \sig{0.381} \\
      $22$ & $1005$ & \sig{0.975} & {\color{forestgreen}$5.96\cdot10^{-7}$} & \sig{1.00} & \sig{0.00983} & \sig{0.0571}  \\
      $34$ & $1485$ & \sig{0.855} & {\color{forestgreen}$5.00\cdot10^{-7}$} & \sig{0.971} & \sig{0.000729} & \sig{0.000128} \\
      $47$ & $2005$ & \sig{0.786} & {\color{forestgreen}$5.00\cdot10^{-7}$} & \sig{0.388} & \sig{0.000932} & {\color{forestgreen}$4.99\cdot10^{-7}$} \\
      $122$ & $5005$ & \sig{0.0131} & {\color{forestgreen}$2.99\cdot10^{-6}$} & {\color{forestgreen}$1.28\cdot10^{-7}$} & \sig{0.131} & {\color{forestgreen}$1.30\cdot10^{-8}$} \\
      $247$ & $10005$ & {\color{forestgreen}$3.46\cdot10^{-6}$} & {\color{forestgreen}$3.96\cdot10^{-5}$} & {\color{forestgreen}$1.86\cdot10^{-9}$} & \sig{0.908} & {\color{forestgreen}$1.57\cdot10^{-7}$} \\
      $372$ & $15005$ & {\color{forestgreen}$7.10\cdot10^{-7}$} & \sig{0.000190} & {\color{forestgreen}$1.86\cdot10^{-9}$} & \sig{0.975} & {\color{forestgreen}$2.86\cdot10^{-7}$} \\
   \end{tabular}
  \end{center}
\end{table}
\begin{table}[h!]
\begin{center}
    \caption{$p$-values Wilcoxon test {\textproc{jas.main}} vs.\ \dots with $L(0) = 250$}
    \label{tab:pl250nslkdd}
    \begin{tabular}{c|c|c|c|c|c|c} 
      $t_{\text{ref}}$ & $L(t_{\text{ref}})$ & {\textproc{jas.basic}} & {\textproc{jas.rand}} & {\textproc{jas.anom}} & {\textproc{jas.uncert}} & {\textproc{ala.main}} \\
      \hline
      $9$ & $610$ & \sig{0.924} & \sig{0.000128} & \sig{0.994} & \sig{0.0790} & \sig{0.516} \\
      $16$ & $890$ & \sig{0.958} & {\color{forestgreen}$2.57\cdot10^{-6}$} & \sig{1.00} & \sig{0.118} & \sig{0.0502}  \\
      $22$ & $1130$ & \sig{0.963} & {\color{forestgreen}$4.99\cdot10^{-7}$} & \sig{1.00} & \sig{0.122} & \sig{0.00530}  \\
      $34$ & $1610$ & \sig{0.915} & {\color{forestgreen}$1.28\cdot10^{-7}$} & \sig{0.999} & {0.388} & {\color{forestgreen}$3.14\cdot10^{-5}$} \\
      $47$ & $2130$ & \sig{0.897} & {\color{forestgreen}$1.02\cdot10^{-7}$} & \sig{0.967} & \sig{0.556} & {\color{forestgreen}$1.38\cdot10^{-6}$} \\
      $122$ & $5130$ & \sig{0.122} & {\color{forestgreen}$4.00\cdot10^{-8}$} & \sig{0.00233} & \sig{0.997} & {\color{forestgreen}$9.31\cdot10^{-9}$} \\
      $247$ & $10130$ & {\color{forestgreen}$9.43\cdot10^{-5}$} & {\color{forestgreen}$4.16\cdot10^{-7}$} & {\color{forestgreen}$8.20\cdot10^{-8}$} & \sig{1.00} & {\color{forestgreen}$8.20\cdot10^{-8}$} \\
      $372$ & $15130$ & {\color{forestgreen}$1.89\cdot10^{-6}$} & {\color{forestgreen}$1.90\cdot10^{-6}$} & {\color{forestgreen}$2.79\cdot10^{-9}$} & \sig{1.00} & {\color{forestgreen}$1.62\cdot10^{-6}$} \\
    \end{tabular}
  \end{center}
\end{table}

To determine whether Jasmine performed significantly better than the other five AL methods, we determined the $p$-value of the test in~\eqref{eq:wilcox} for each method $m = 2, \dots, 6$ and for different values of $t_{\text{ref}}$. Table~\ref{tab:pl125nslkdd} and~\ref{tab:pl250nslkdd} show the results for the experiments with $L(0) = 125$ and $L(0) = 250$, respectively. A green value indicates that Jasmine ({\textproc{jas.main}}) performed significantly better than the method in the corresponding column for the labeled set size $L(t_{\text{ref}})$. A red value, however, means that Jasmine performed significantly worse (by interchanging the sides of~\eqref{eq:wilcox}). A black value means that the test was indecisive and could not conclude whether Jasmine was better or worse.

\subsubsection{$\alpha$-dynamic updating}
\begin{figure}[h!]
\centering
\begin{subfigure}{.5\textwidth}
 \centering
 \includegraphics[width=0.95\linewidth]{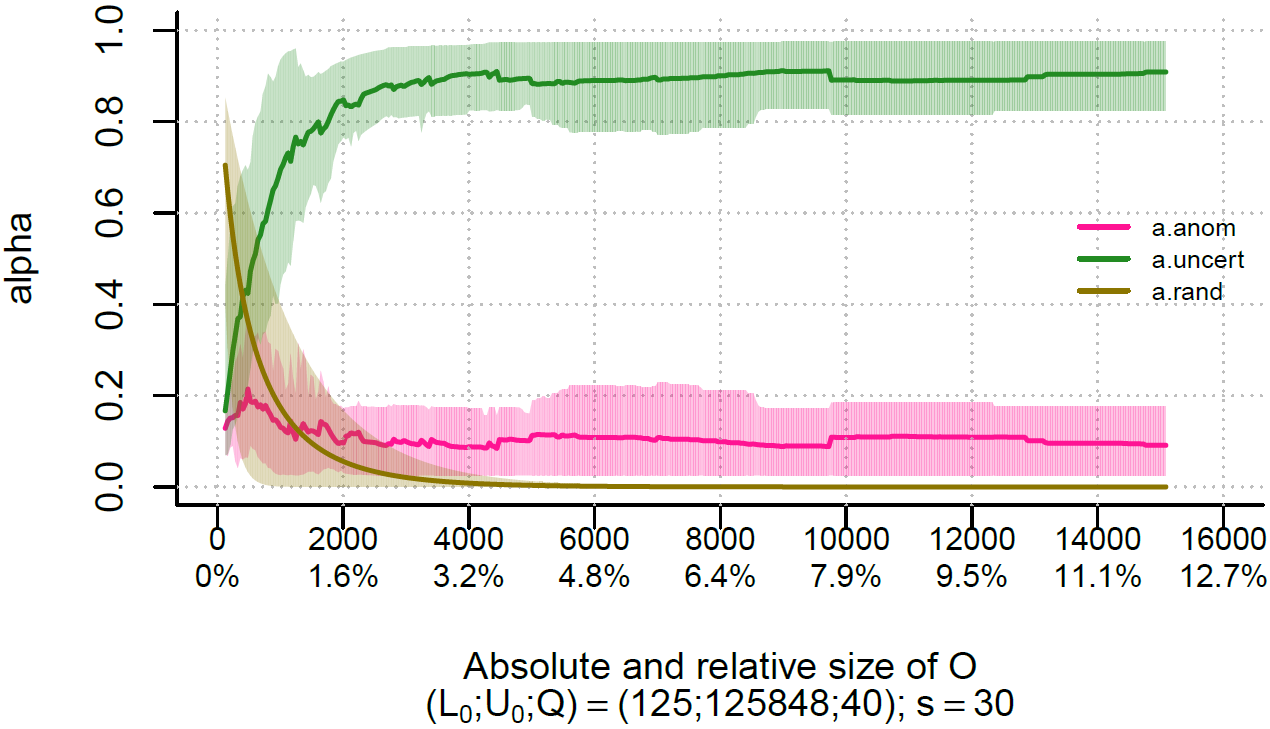}
 \caption{$L(0) = 125$}
 \label{fig:al125nslkdd}
 \end{subfigure}%
 \begin{subfigure}{.5\textwidth}
 \centering
 \includegraphics[width=0.95\linewidth]{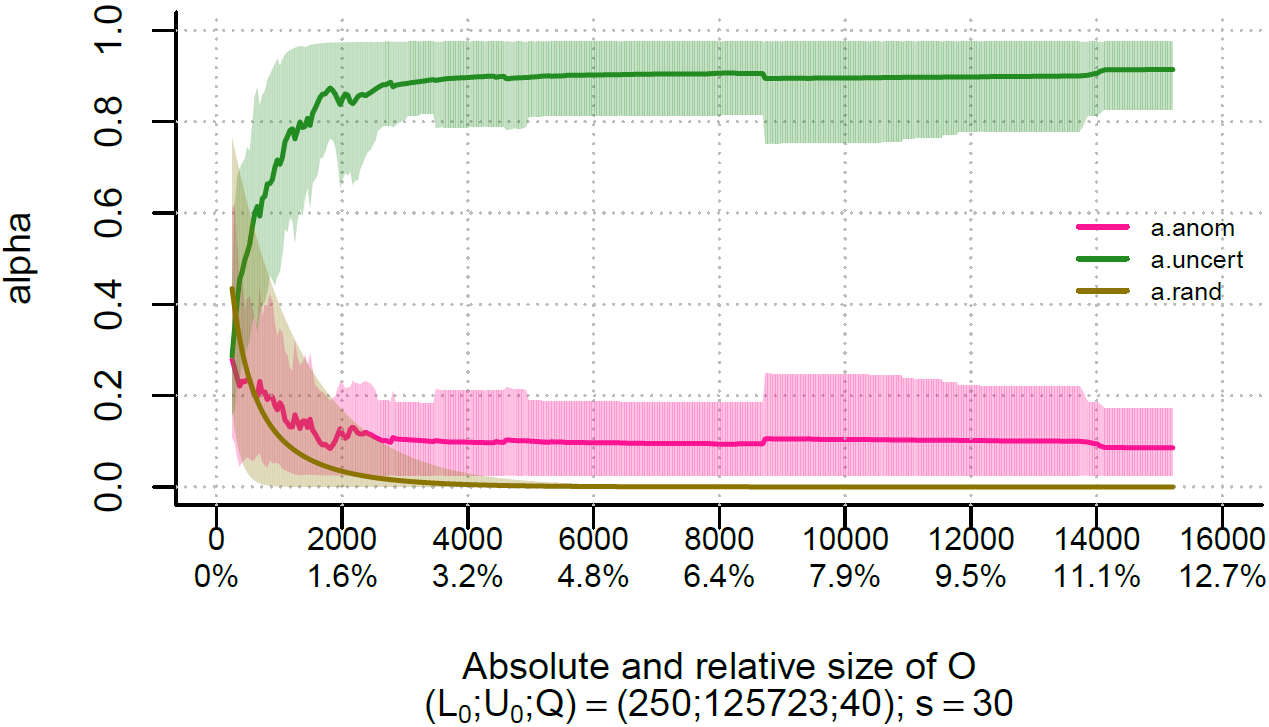}
 \caption{$L(0) = 250$}
 \label{fig:al250nslkdd}
\end{subfigure}
\caption{Progress of query fractions for different initial sizes.}
\label{fig:anslkdd}
\end{figure}
Finally, Figure~\ref{fig:anslkdd} presents the $\alpha$-dynamic updating procedure of Jasmine for $L(0) = 125$ and $L(0) = 250$. Similar to the figure with the learning curves, every simulation resulted in a $\alpha$-dynamic curve for each of the query fractions $\alpha_a(\cdot)$ ({\textproc{a.anom}}), $\alpha_z(\cdot)$ ({\textproc{a.uncert}}) and $\alpha_r(\cdot)$ ({\textproc{a.rand}}). We took the average of the query fractions to obtain the average $\alpha$-dynamic curves. A shaded region indicates the interval in which $80\%$ of the observed fractions with matching color resided, and therefore, shows the spread of the values. Since $S = 30$ simulations were performed, each region contains the 24 `middle' fraction values. 

\subsubsection{Implication of Results on NSL-KDD}
The first thing that we observe in Figure~\ref{fig:f1nslkdd} is that the average $F_1$ score rapidly increased in the first iterations for the Jasmine ({\textproc{jas.main}}), {\textproc{jas.anom}}, {\textproc{jas.uncert}} and {\textproc{jas.basic}} procedures. This means that by using the corresponding query approaches, valuable unlabeled observations were queried to the oracle, because the GBMs were able to make better predictions on the evaluation set NSL-KDD-$\mathcal{E}$. This increase is most notable for \textproc{jas.anom} and then especially for $L(0) = 250$. This makes sense, because NSL-KDD-$\mathcal{E}$ contains new anomalous attack scenarios that are not found in the train set. Remarkably, the performance of \textproc{jas.anom} went higher than the final (average) $F_1$ score. For several iteration steps, also the other three methods obtained scores higher than the final performance for both settings of $L(0)$. This means that a carefully constructed smaller dataset led to better predictions on the evaluation set than the complete train set did. 

Even though \textproc{jas.anom} performed better than the other methods at the start of the iterations, its effectiveness decreased when more anomalous observations were added to the labeled set. 
The decrease of effectiveness is also visible in Table~\ref{tab:pl125nslkdd} and~\ref{tab:pl250nslkdd}: for small values of $t_{\text{ref}}$ querying only anomalies performed better than Jasmine, but later Jasmine obtained significantly better results.
It could be that the labeled set became more and more abnormal when more anomalous observations were added, resulting in impaired training of the GBM classifier in later iterations after it had improved before.

Furthermore, it is clear that Jasmine performed better than only querying uncertainties, as the $p$-values show. It appears that also querying anomalies is important. However, when time progresses, its performance is not significantly better anymore and eventually becomes significantly worse. It should be stressed, though, that the learning curves show that both \textproc{jas.main} and \textproc{jas.uncert} have converged to the final score and only differ a little.

Only querying random observations, as done by \textproc{jas.rand}, performed significantly worse than Jasmine for all reference iterations. This shows that specifically choosing anomalous or uncertain observations works better than only querying random observations.

Also interesting to note is that Jasmine was on par with or worse than \textproc{jas.basic} at the start of the iteration procedure. However, when the size of $\mathcal{L}(\cdot)$ grew, Jasmine became significantly better, as the $p$-values in both Table~\ref{tab:pl125nslkdd} and~\ref{tab:pl250nslkdd} show. This means that dynamically adjusting the query balance in a later stadium has an advantage over querying anomalies and uncertainties in a fixed 50/50 fashion. Combining this with the progress of the query fractions in Figure~\ref{fig:anslkdd} and with the fact that \textproc{jas.anom}'s performance worsens over time shows that Jasmine found the right balance between querying uncertainties and anomalies. 
Nonetheless, since anomalies appeared to be better at the start, it is curious why Jasmine did not query more anomalies in that stage. Hence, the information metrics as defined in~\eqref{eq:anomdelta} and~\eqref{eq:uncertdelta} could have a preference for uncertain over anomalous observations.

Lastly, Jasmine performed fairly quickly significantly better than ALADIN as the $p$-values show. This is partly due to the simpler ML techniques in the latter, as it took Jasmine less effort to obtain better results than ALADIN than it took to perform better than \textproc{jas.basic}. 

In general, there do not seem to be large differences between the experiments with $L(0) = 125$ and with $L(0) = 250$. However, the $\alpha$-dynamic curves in Figure~\ref{fig:anslkdd} show that the average initial random query fraction $\alpha_r(0)$ was noticeably bigger for $L(0) = 125$ than for $L(0) = 250$, since the brown curve (\textproc{a.rand}) starts higher in the former. This makes sense, because $\mathcal{L}(0)$ was randomly constructed, and hence, it became less essential to query random instances when we chose $L(0)$ larger.

\subsection{Results on NSL-KDD-rand}
\label{subsec:nslkddrand}

\subsubsection{Learning curves}
\begin{figure}[h!]
\centering
\begin{subfigure}[b]{.65\textwidth}
 \centering
 \includegraphics[width=1\textwidth]{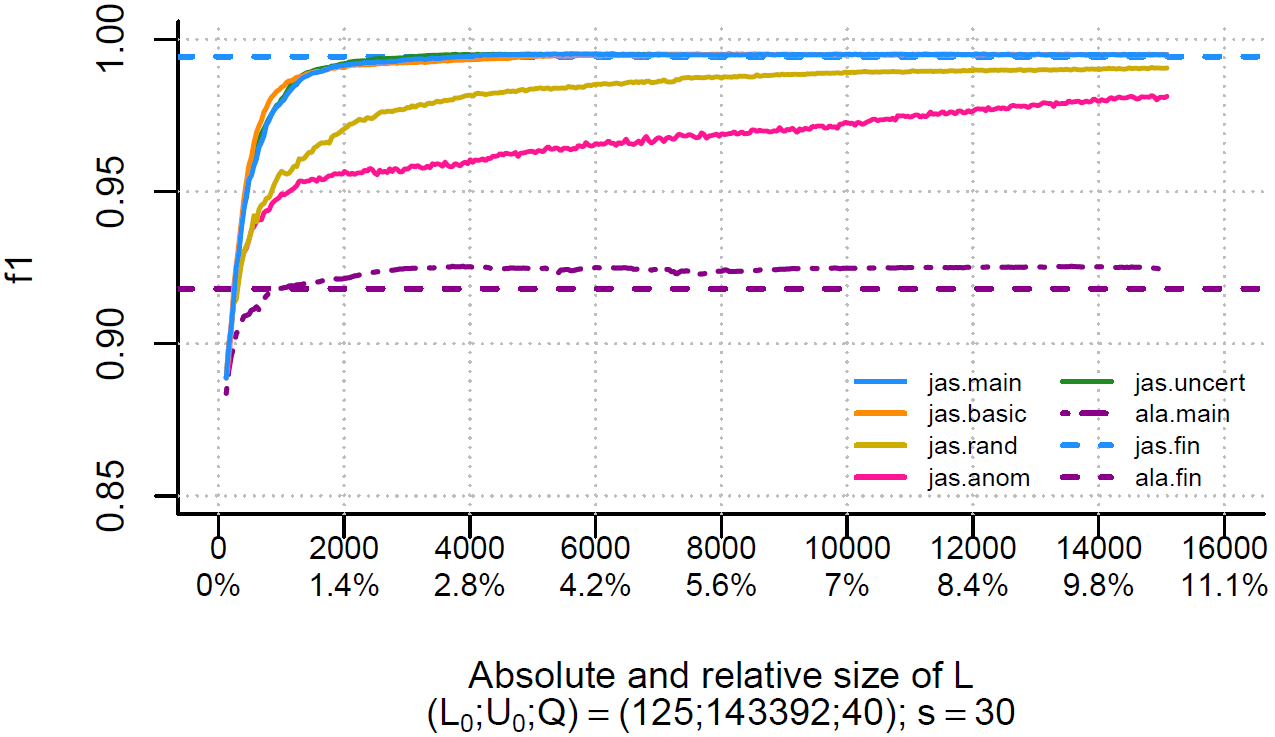}
 \caption{$L(0) = 125$}
 \label{fig:f1l125nslkddrand}
\end{subfigure}

\begin{subfigure}[b]{.65\textwidth}
 \centering
 \includegraphics[width=1\textwidth]{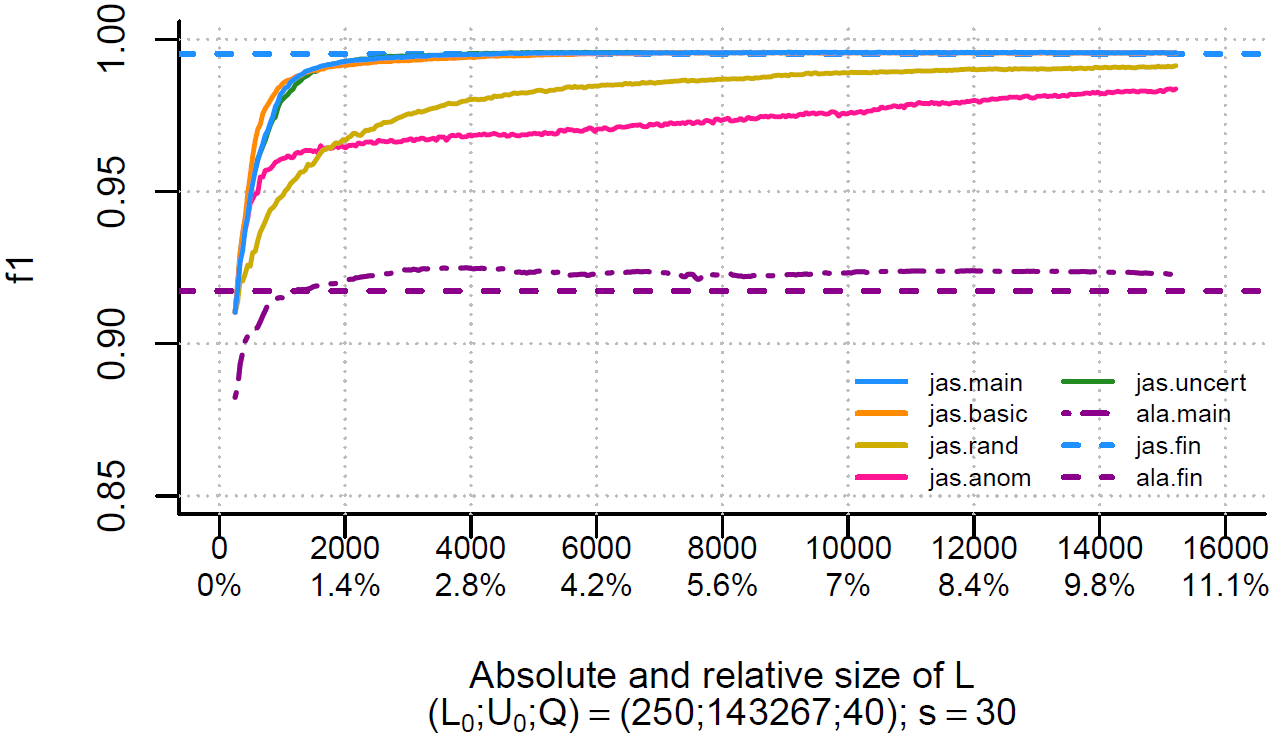}
 \caption{$L(0) = 250$}
 \label{fig:f1l250nslkddrand}
\end{subfigure}
\caption{Learning curves on NSL-KDD with random evaluation set for different $L(0)$.}
\label{fig:f1nslkddrand}
\end{figure}
Similar to the results presented in Section~\ref{subsec:nslkdd}, Figure~\ref{fig:f1nslkddrand} presents the average learning curves for each of the six AL methods and two constant reference lines. In contrast to the results on the NSL-KDD dataset, the evaluation set $\mathcal{E}$ was chosen anew for each simulation. More specifically, at the start of each repetition the complete NSL-KDD-rand dataset was randomly partitioned in $\mathcal{L}(0)$, $\mathcal{U}(0)$ and $\mathcal{E}$. 
The blue dashed line ({\textproc{jas.fin}}) is the average final performance of the simulation-specific GBMs on their corresponding evaluation sets. This metric was $\overline{F_1} \approx 0.994$ 
for Figure~\ref{fig:f1l125nslkddrand} and $\overline{F_1} \approx 0.995$ 
for Figure~\ref{fig:f1l250nslkddrand}. This time, the average final performance of ALADIN represented by the purple dashed line ({\textproc{ala.fin}}) was no longer necessarily constant, but $\overline{F_1} \approx 0.918$ for initial set size $L(0) = 125$ and $\overline{F_1} \approx 0.917$ for $L(0) = 250$. 
Lastly, the expected performance of the best dummy classifier is not visible in the figures, since it was $\overline{F_1} \approx 0.650$ 
for $L(0) = 125$ and $\overline{F_1} \approx 0.651$ 
for $L(0) = 250$.

\subsubsection{Statistical tests}
\begin{table}[h!]
  \begin{center}
    \caption{$p$-values Wilcoxon test {\textproc{jas.main}} vs.\ \dots with $L(0) = 125$}
    \label{tab:pl125nslkddrand}
   \begin{tabular}{c|c|c|c|c|c|c} 
      $t_{\text{ref}}$ & $L(t_{\text{ref}})$ & {\textproc{jas.basic}} & {\textproc{jas.rand}} & {\textproc{jas.anom}} & {\textproc{jas.uncert}} & {\textproc{ala.main}} \\
      \hline
      $9$ & $485$ & \sig{0.798} & \sig{0.000365} & \sig{0.0192} & \sig{0.657} & \fg{5.96}{-7} \\
      $16$ & $765$ & \sig{0.886} & \fg{5.96}{-7} & \sig{0.000333} & \sig{0.657} & \fg{1.86}{-9} \\
      $22$ & $1005$ & \sig{0.904} & \fg{1.30}{-8} & \fg{1.52}{-5} & \sig{0.642} & \fg{1.86}{-9}  \\
      $34$ & $1485$ & \sig{0.894} & \fg{9.31}{-10} & \fg{8.20}{-8} & \sig{0.672} & \fg{9.31}{-10} \\
      $47$ & $2005$ & \sig{0.831} & \fg{9.31}{-10} & \fg{9.31}{-10} & \sig{0.708} & \fg{9.31}{-10} \\
      $122$ & $5005$ & \sig{0.492} & \fg{9.31}{-10} & \fg{9.31}{-10} & \sig{0.836} & \fg{9.31}{-10} \\
      $247$ & $10005$ & \sig{0.428} & \fg{9.31}{-10} & \fg{9.31}{-10} & \sig{0.846} & \fg{9.31}{-10} \\
      $372$ & $15005$ & \sig{0.365} & \fg{9.31}{-10} & \fg{9.31}{-10} & \sig{0.841} & \fg{9.31}{-10} \\
   \end{tabular}
  \end{center}
\end{table}
\begin{table}[h!]
  \begin{center}
    \caption{$p$-values Wilcoxon test {\textproc{jas.main}} vs.\ \dots with $L(0) = 250$}
    \label{tab:pl250nslkddrand}
   \begin{tabular}{c|c|c|c|c|c|c} 
      $t_{\text{ref}}$ & $L(t_{\text{ref}})$ & {\textproc{jas.basic}} & {\textproc{jas.rand}} & {\textproc{jas.anom}} & {\textproc{jas.uncert}} & {\textproc{ala.main}} \\
      \hline
      $9$ & $610$ & \sig{0.970} & \fg{4.16}{-7} & \sig{0.313} & \sig{0.930} & \fg{2.79}{-9} \\
      $16$ & $890$ & \sig{0.985} & \fg{1.28}{-7} & \sig{0.0261} & \sig{0.701} & \fg{1.86}{-9} \\
      $22$ & $1130$ & \sig{0.975} & \fg{1.77}{-8} & \fg{8.59}{-4} & \sig{0.548} & \fg{9.31}{-10}  \\
      $34$ & $1610$ & \sig{0.963} & \fg{9.31}{-10} & \fg{5.96}{-7} & \sig{0.516} & \fg{9.31}{-10} \\
      $47$ & $2130$ & \sig{0.945} & \fg{9.31}{-10} & \fg{9.31}{-10} & \sig{0.564} & \fg{9.31}{-10} \\
      $122$ & $5130$ & \sig{0.635} & \fg{9.31}{-10} & \fg{9.31}{-10} & \sig{0.650} & \fg{9.31}{-10} \\
      $247$ & $10130$ & \sig{0.444} & \fg{9.31}{-10} & \fg{9.31}{-10} & \sig{0.687} & \fg{9.31}{-10} \\
      $372$ & $15130$ & \sig{0.444} & \fg{9.31}{-10} & \fg{9.31}{-10} & \sig{0.650} & \fg{9.31}{-10} \\
   \end{tabular}
  \end{center}
\end{table}

Table~\ref{tab:pl125nslkddrand} and~\ref{tab:pl250nslkddrand} show the $p$-values of the Wilcoxon signed-rank test with null hypothesis as given in~\eqref{eq:wilcox} for different values of $t_{\text{ref}}$. 

\subsubsection{$\alpha$-dynamic updating}
\begin{figure}[h!]
\centering
\begin{subfigure}{.5\textwidth}
 \centering
 \includegraphics[width=0.95\textwidth]{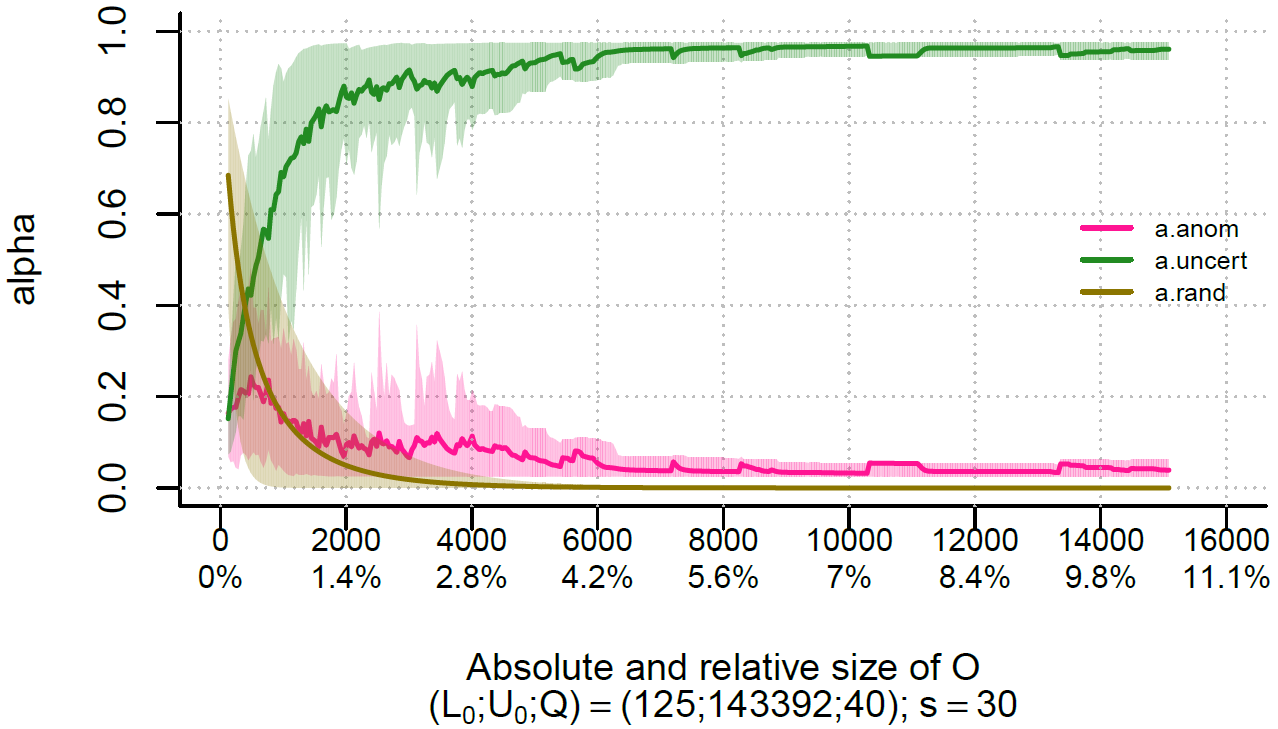}
 \caption{$L(0) = 125$}
 \label{fig:al125nslkddrand}
\end{subfigure}%
\begin{subfigure}{.5\textwidth}
 \centering
 \includegraphics[width=0.95\textwidth]{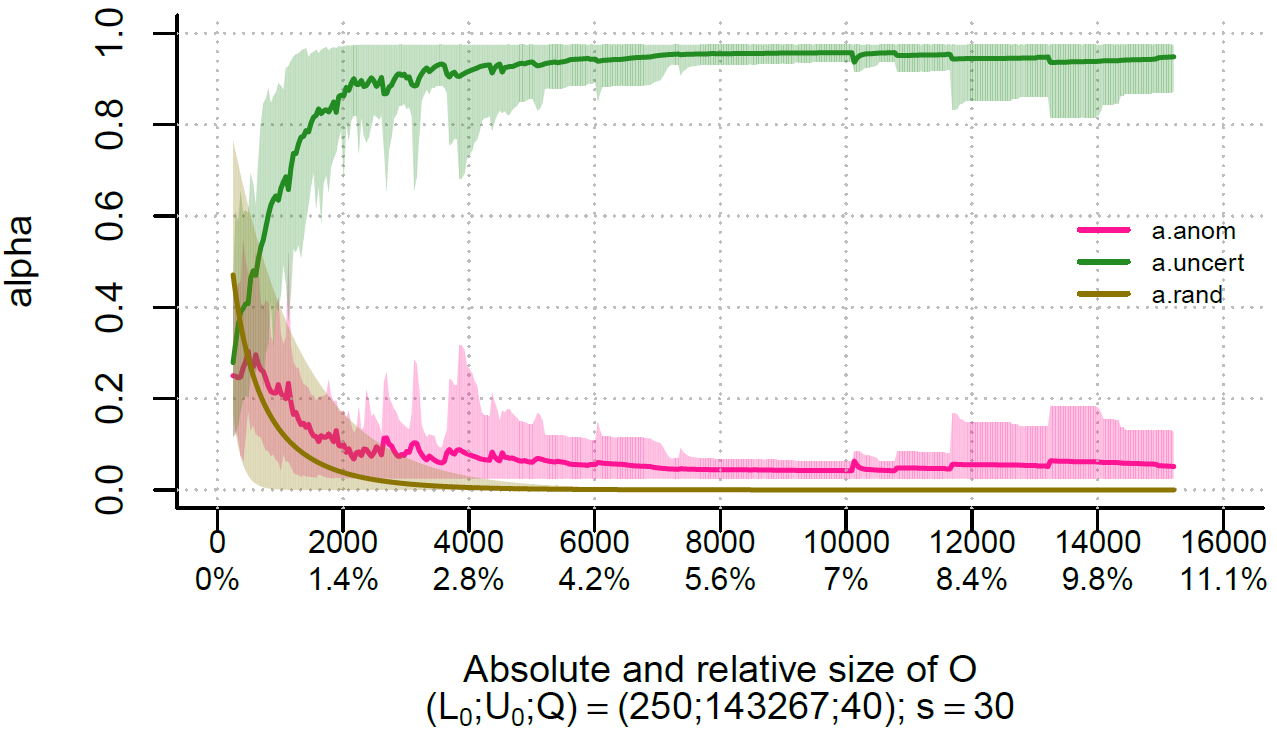}
 \caption{$L(0) = 250$}
 \label{fig:al250nslkddrand}
\end{subfigure}
\caption{Progress of query fractions for different initial sizes.}
\label{fig:anslkddrand}
\end{figure}

The dynamics of the $\alpha$-updating procedure of Jasmine are shown in Figure~\ref{fig:anslkddrand} for $L(0) = 125$ and $L(0) = 250$. As described before, the curves represent the average query fractions $\alpha_a(\cdot)$ ({\textproc{a.anom}}), $\alpha_z(\cdot)$ ({\textproc{a.uncert}}) and $\alpha_r(\cdot)$ ({\textproc{a.rand}}) throughout the iteration process. 

\subsubsection{Implication of Results on NSL-KDD-rand}
At first glance, the results on the randomly selected evaluation sets of NSL-KDD-rand are much better for every considered AL method. 
This seems reasonable, because the fixed evaluation set for the NSL-KDD data contains unseen cyberattacks on which the classifier could not train. In the case of NSL-KDD-rand, the evaluation set was randomly chosen from the complete dataset, so we expected it to have the same structure as the train set, making it easier for the classifier to learn. This was specifically the case for the GBM, because it rapidly obtained an almost perfect $F_1$ score on the evaluation set.

The figures also show that the three learning curves for Jasmine, \textproc{jas.basic} and \textproc{jas.uncert} increased in a similar fashion at the start of the procedure. This was not the case for \textproc{jas.anom}, indicating that it was mostly important to query uncertain observations. This was also reflected in the $p$-values, as shown by Table~\ref{tab:pl125nslkddrand} and~\ref{tab:pl250nslkddrand}. Jasmine quickly performed significantly better than \textproc{jas.anom}, but it had more difficulty in obtaining significantly better results than \textproc{jas.basic} and \textproc{jas.uncert}. The latter was even significantly better than Jasmine. However, it should be noted that the $F_1$ scores were already near perfect for these three query approaches. 
Moreover, the dynamics of the query fractions in Figure~\ref{fig:anslkddrand} show that there was a clear preference for querying more uncertainties than anomalous observations. 

Furthermore, the learning curves show that Jasmine obtained better results than \textproc{jas.rand} and ALADIN. Both Table~\ref{tab:pl125nslkddrand} and~\ref{tab:pl250nslkddrand} indicate that Jasmine was significantly better for all considered reference values. 


\subsection{Results on UNSW-NB15}

\subsubsection{Learning curves}
\begin{figure}[h!]
\centering
\begin{subfigure}[b]{.65\textwidth}
 \centering
 \includegraphics[width=1\textwidth]{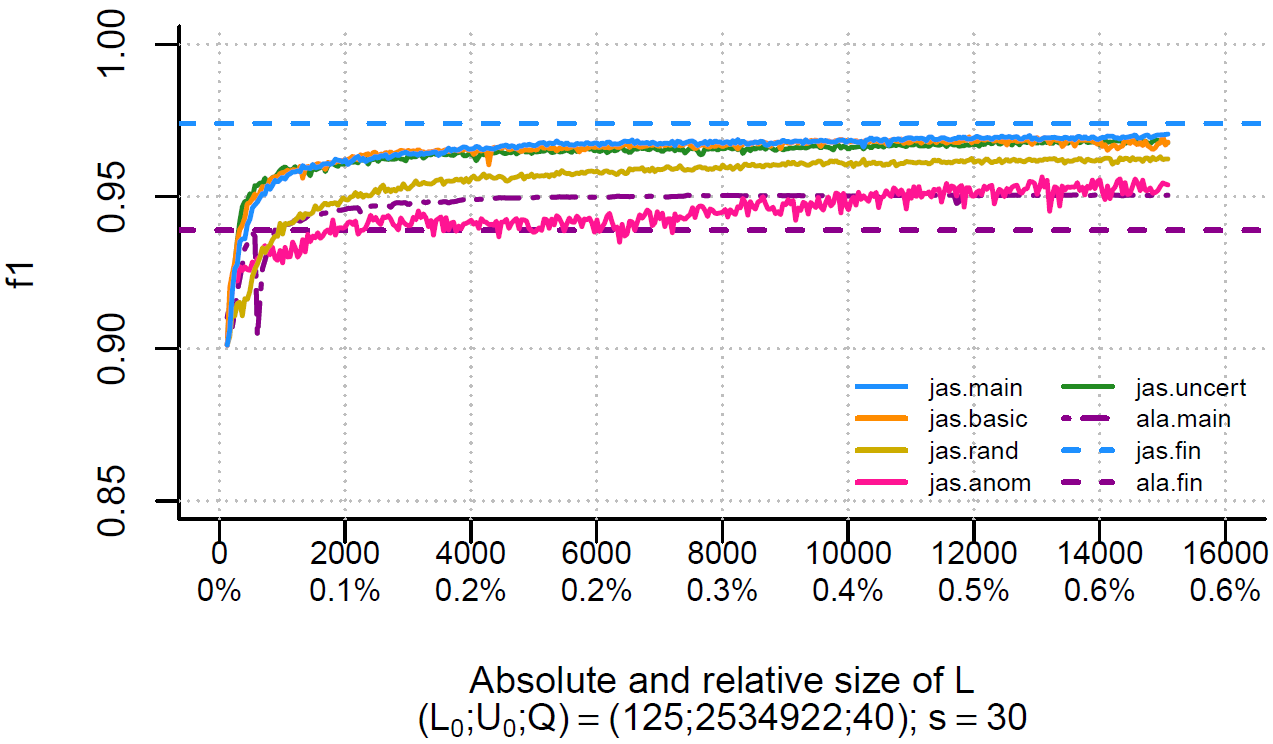}
 \caption{$L(0) = 125$}
 \label{fig:f1l125unsw}
\end{subfigure}

\begin{subfigure}[b]{.65\textwidth}
 \centering
 \includegraphics[width=1\textwidth]{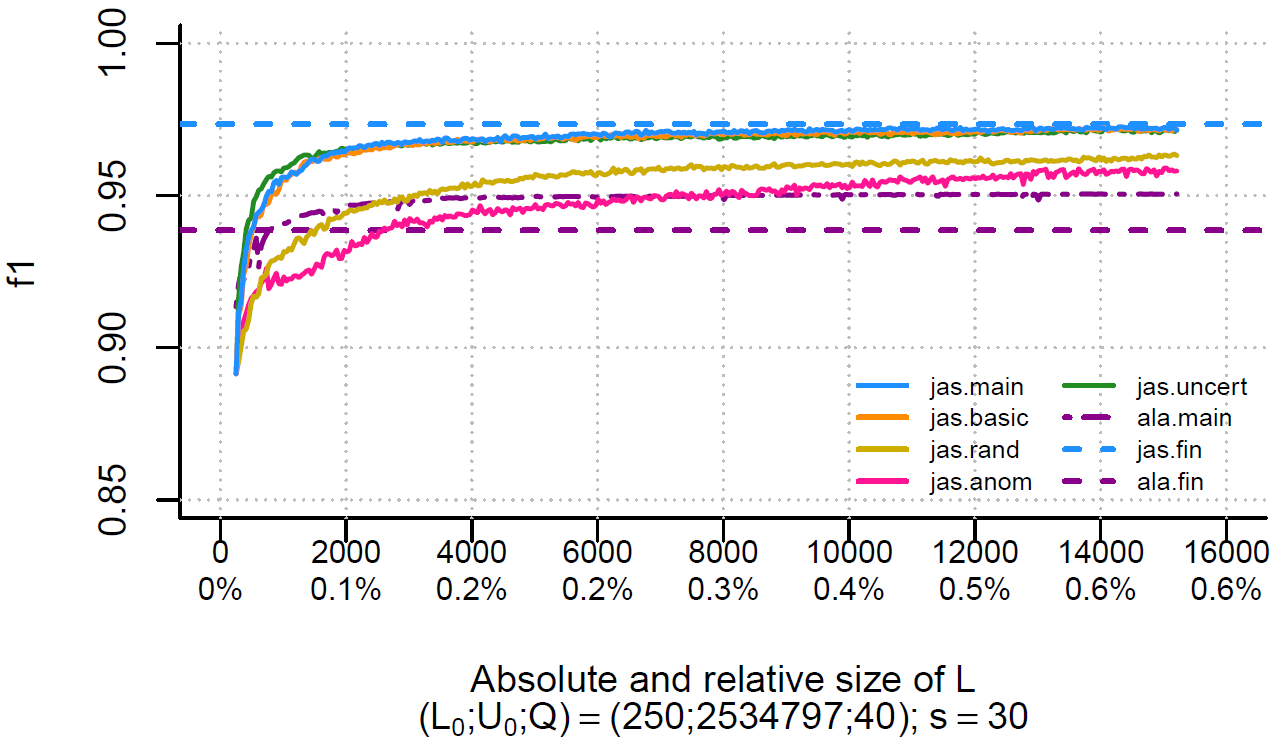}
 \caption{$L(0) = 250$}
 \label{fig:f1l250unsw}
\end{subfigure}
\caption{Learning curves on UNSW-NB15 with random evaluation set for different $L(0)$.}
\label{fig:f1unsw}
\end{figure}

Figure~\ref{fig:f1unsw} shows the average learning curves for every one of the six AL methods and two constant reference lines. The evaluation set $\mathcal{E}$ was chosen anew for each simulation, equivalent to what we discussed in Section~\ref{subsec:nslkddrand}. The blue dashed line ({\textproc{jas.fin}}) is the average final performance of the simulation-specific GBMs on their corresponding evaluation sets. For Figure~\ref{fig:f1l125unsw}, this metric was $\overline{F_1} \approx 0.974$, while it was $\overline{F_1} \approx 0.973$ for Figure~\ref{fig:f1l250unsw}. The average final performance of ALADIN indicated by the purple dashed line ({\textproc{ala.fin}}) was $\overline{F_1} \approx 0.939$ for both initial set sizes. Lastly, the figures do no show the expected performance of the best dummy classifier, since it was $\overline{F_1} \approx 0.225$ for $L(0) = 125$ and $\overline{F_1} \approx 0.222$ for $L(0) = 250$. These remarkably lower baseline values are because relatively far fewer positive observations are present in UNSW-NB15 compared to NSL-KDD.

\subsubsection{Statistical tests}
\begin{table}[h!]
  \begin{center}
    \caption{$p$-values Wilcoxon test {\textproc{jas.main}} vs.\ \dots with $L(0) = 125$}
    \label{tab:pl125unsw}
   \begin{tabular}{c|c|c|c|c|c|c} 
      $t_{\text{ref}}$ & $L(t_{\text{ref}})$ & {\textproc{jas.basic}} & {\textproc{jas.rand}} & {\textproc{jas.anom}} & {\textproc{jas.uncert}} & {\textproc{ala.main}} \\
      \hline
      $9$ & $485$ & \sig{0.994} & \fg{4.95}{-5} & \sig{0.0213} & \sig{0.994} & \sig{0.0701} \\
      $16$ & $765$ & \sig{0.990} & \fg{1.90}{-6} & \fg{3.96}{-5} & \sig{0.997} & \fg{6.87}{-5} \\
      $22$ & $1005$ & \sig{0.985} & \fg{9.96}{-7} & \fg{4.00}{-6} & \sig{0.992} & \fg{3.46}{-6}  \\
      $34$ & $1485$ & \sig{0.965} & \fg{1.93}{-7} & \fg{1.28}{-7} & \sig{0.971} & \fg{1.30}{-8} \\
      $47$ & $2005$ & \sig{0.918} & \fg{4.00}{-8} & \fg{6.52}{-9} & \sig{0.897} & \fg{9.31}{-9} \\
      $122$ & $5005$ & \sig{0.815} & \fg{9.31}{-9} & \fg{1.86}{-9} & \sig{0.612} & \fg{2.79}{-9} \\
      $247$ & $10005$ & \sig{0.271} & \fg{5.12}{-8} & \fg{1.86}{-9} & \sig{0.191} & \fg{1.86}{-9} \\
      $372$ & $15005$ & \sig{0.122} & \fg{1.28}{-7} & \fg{1.86}{-9} & \sig{0.131} & \fg{1.86}{-9} \\
   \end{tabular}
  \end{center}
\end{table}
\begin{table}[h!]
  \begin{center}
    \caption{$p$-values Wilcoxon test {\textproc{jas.main}} vs.\ \dots with $L(0) = 250$}
    \label{tab:pl250unsw}
   \begin{tabular}{c|c|c|c|c|c|c} 
      $t_{\text{ref}}$ & $L(t_{\text{ref}})$ & {\textproc{jas.basic}} & {\textproc{jas.rand}} & {\textproc{jas.anom}} & {\textproc{jas.uncert}} & {\textproc{ala.main}} \\
      \hline
      $9$ & $610$ & \sig{0.715} & \fg{3.46}{-7} & \fg{7.99}{-6} & \sig{0.989} & \sig{0.149} \\
      $16$ & $890$ & \sig{0.735} & \fg{5.96}{-7} & \fg{1.90}{-6} & \sig{0.998} & \sig{0.00128} \\
      $22$ & $1130$ & \sig{0.761} & \fg{4.16}{-7} & \fg{2.36}{-7} & \sig{0.997} & \sig{0.000230}  \\
      $34$ & $1610$ & \sig{0.650} & \fg{9.31}{-9} & \fg{1.77}{-8} & \sig{0.998} & \fg{1.52}{-5} \\
      $47$ & $2130$ & \sig{0.350} & \fg{2.79}{-9} & \fg{9.31}{-10} & \sig{0.990} & \fg{2.57}{-6} \\
      $122$ & $5130$ & \sig{0.185} & \fg{9.31}{-10} & \fg{9.31}{-10}& \sig{0.786} & \fg{9.31}{-10} \\
      $247$ & $10130$ & \sig{0.0481} & \fg{9.31}{-10}& \fg{9.31}{-10} & \sig{0.306} & \fg{9.31}{-10} \\
      $372$ & $15130$ & \sig{0.0275} & \fg{9.31}{-10} & \fg{9.31}{-10} & \sig{0.180} & \fg{9.31}{-10} \\
   \end{tabular}
  \end{center}
\end{table}

Table~\ref{tab:pl125unsw} and~\ref{tab:pl250unsw} present the $p$-values of the Wilcoxon signed-rank test with null hypothesis as given in~\eqref{eq:wilcox} for different values of $t_{\text{ref}}$.

\subsubsection{$\alpha$-dynamic updating}
\begin{figure}[h!]
\centering
\begin{subfigure}{.5\textwidth}
 \centering
 \includegraphics[width=0.95\textwidth]{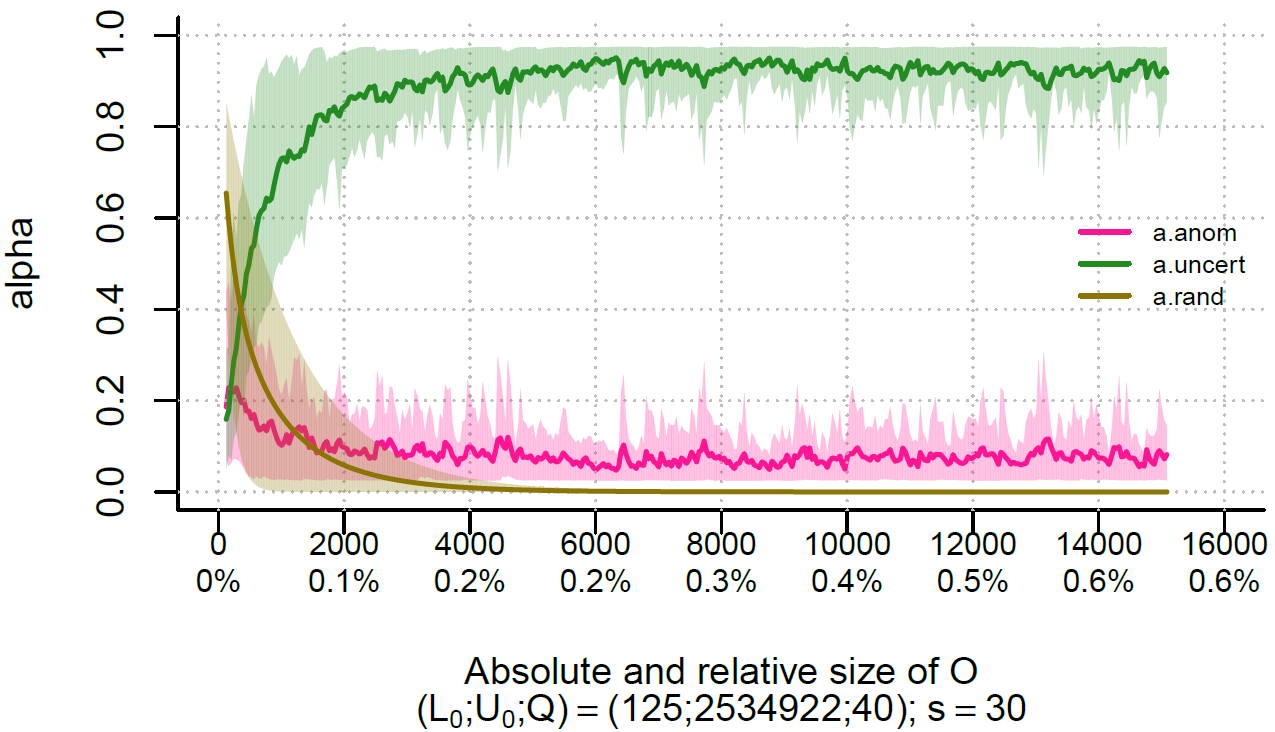}
 \caption{$L(0) = 125$}
 \label{fig:al125unsw}
\end{subfigure}%
\begin{subfigure}{.5\textwidth}
 \centering
 \includegraphics[width=0.95\textwidth]{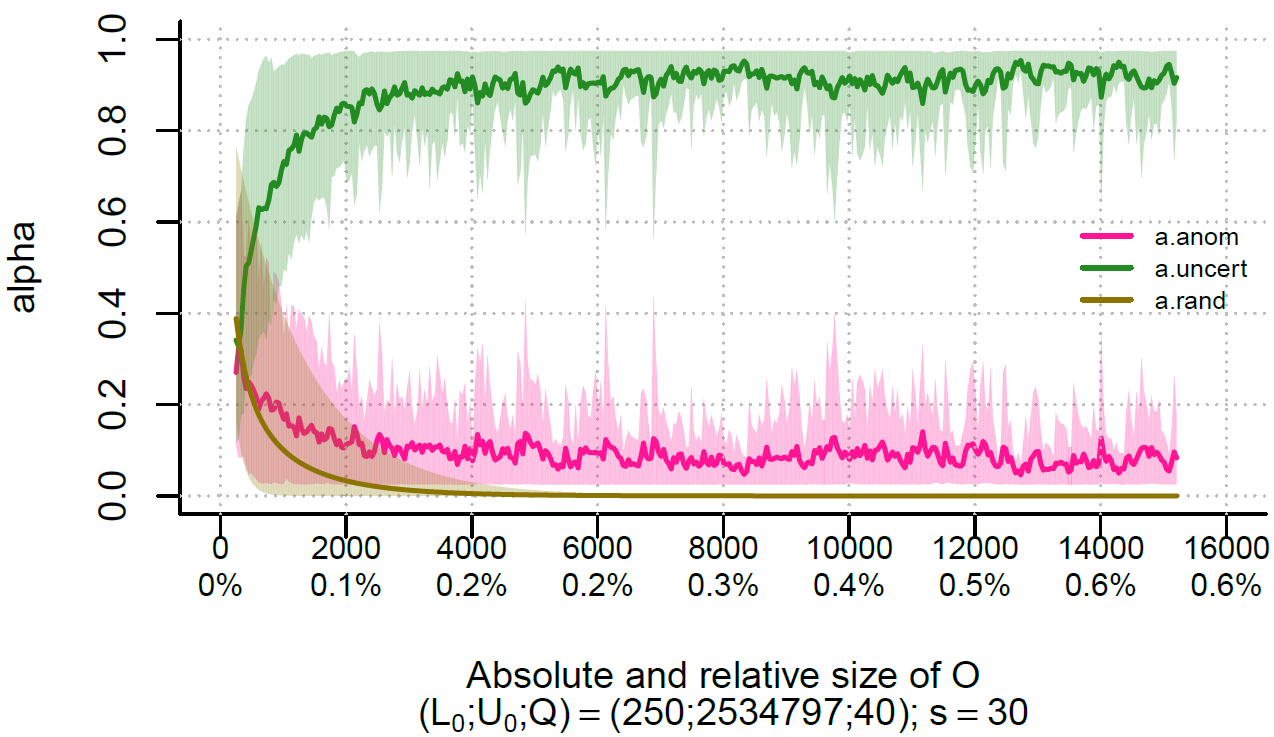}
 \caption{$L(0) = 250$}
 \label{fig:al250unsw}
\end{subfigure}
\caption{Progress of query fractions for different initial size $L(0)$.}
\label{fig:aunsw}
\end{figure}

Figure~\ref{fig:aunsw} shows the $\alpha$-dynamic updating procedure of Jasmine for $L(0) = 125$ and $L(0) = 250$. Equivalent to the results in Section~\ref{subsec:nslkdd} and~\ref{subsec:nslkddrand}, the curves indicate the average query fractions $\alpha_a(\cdot)$ ({\textproc{a.anom}}), $\alpha_z(\cdot)$ ({\textproc{a.uncert}}) and $\alpha_r(\cdot)$ ({\textproc{a.rand}}) throughout the iteration process.

\subsubsection{Implication of Results on UNSW-NB15}
There are no striking differences between the two plots in Figure~\ref{fig:f1unsw}. However, it seems that the average learning curves for initial set size $L(0) = 125$ are a bit more shaky. Since $L(0)$ is the only different global parameter between the two plots, the change in behavior is presumably due to how the GBM parameters and Jasmine-specific parameters were tuned. It probably also had to do with the imbalance of this dataset. Approximately 12.6\% of the data is related to cyberattacks, so on average about $16$ observations were malicious in $\mathcal{L}(0)$. For tuning purposes, this initial set was split in training, validation and test sets, making it possible that only one malicious observation ended up in any of those sets. Consequently, the tuned parameters possibly led to less stable behavior of the GBM and the $\alpha$-dynamic update procedure. Figure~\ref{fig:aunsw} supports the latter, since more spikes are apparent in the confidence regions than in the update curves for the other two datasets. However, this is the case for both initial set sizes, making it more likely that it is related to the tuning of the GBM. Hence, balancing techniques of the training data such as under- or oversampling could be considered to improve stability.

Furthermore, the results for the learning curves in Figure~\ref{fig:f1unsw} are similar to those obtained for NSL-KDD-rand. This is also true for the $p$-values presented in Table~\ref{tab:pl125unsw} and~\ref{tab:pl250unsw} and the dynamics of the average $\alpha$-update curves shown in Figure~\ref{fig:aunsw}.

\section{Discussion}
\label{sec:discussion}

In this final section of the paper, we firstly draw conclusions based on the implications of the results discussed in Section~\ref{sec:results}. Secondly, we consider further directions for AL in the field of network intrusion detection.

\subsection{Conclusion}
\label{subsec:conclusion}

The goal of this research was to propose our hybrid Active Learning method Jasmine for network intrusion detection. It consists of $\alpha$-dynamic querying, which means Jasmine is able to dynamically adjust the balance between querying anomalous, uncertain and random observations. Consequently, only the potentially most  interesting observations are presented to the human expert. This sets our method apart from other AL approaches that have a static query function. 
In this paper, we firstly formulated the mathematical foundation of Jasmine. Then, we described how the experiments were set up to determine whether our method achieved good results. The first dataset we chose for this analysis was the commonly used NSL-KDD data. Also, we restructured this data to obtain the second dataset NSL-KDD-rand with other properties. Additionally, we also chose the more recent and realistic UNSW-NB15 data.

On all three datasets, Jasmine performed significantly better than ALADIN, the benchmark AL method used in this research. 
This means for practitioners that it is beneficial to choose Jasmine over ALADIN for their AL problems. However, we should note that Jasmine needs preprocessing time, which ALADIN and other static AL methods do not need. This is because of GBM tuning and Jasmine tuning.

Furthermore, we observed that Jasmine performed more robustly with respect to the datasets. We compared the characteristic $\alpha$-dynamic query function of Jasmine with other query functions: querying only anomalies, only uncertainties, only random observations and querying anomalous and uncertain observations in a fixed 50/50 fashion. We noticed that Jasmine did not always outperform the other approaches, but its performance was less influenced by the considered data, and hence, more robust. On the NSL-KDD dataset, only querying anomalies performed better when the labeling process had just started, but Jasmine took over in the long run. On the NSL-KDD-rand and UNSW-NB15 data, only querying uncertainties or querying in a 50/50 fashion reigned supreme, but Jasmine followed closely. Only querying anomalies was clearly a bad strategy for these two datasets. These findings suggest that Jasmine is better able to adapt to the provided labeled dataset $\mathcal{L}(\cdot)$. This is particularly interesting in the face of concept drift: NIDSs operate in high non-stationary contexts, which makes it wise to consider an AL method that is able to adjust its query approach dynamically.

However, there is room for improvement in the way $\alpha$-dynamic querying is performed here. The results on NSL-KDD show that at first querying more anomalous than uncertain observations was beneficial, but this was not reflected in the way that the query fractions were updated. The updating procedure seems to have a bias towards querying uncertainties.

\subsection{Future Work}

Our first suggestion for further research is to reconsider the $\alpha$-dynamic query process such that the bias towards selecting uncertain observations is eliminated.

Another suggestion is to add unlabeled observations about which the classifier has a high prediction certainty to the labeled set without asking the oracle for its label. Consequently, the labeled set increases much more during an iteration, while the human expert does not have to label more observations. This reduces labeling time drastically and possibly leads to better predictions in an earlier stage of the AL process. 
Moreover, when more labels become available, the preprocessing steps can be repeated (in a less extensive version) to allow for the GBM to better adjust to the changing train set by retuning its hyperparameters. Also, some Jasmine-specific parameters could be retuned during the process.

Our last suggestion is to consider human uncertainty in the AL method, since it is not always clear whether a connection is benign or malicious. This can be done by asking the oracle for their confidence in each label that they provide or by incorporating a general probability. 

These suggestions would increase the deployment efficiency of Jasmine even more. Therefore, we would like to apply our method in a practical setting to see how it performs.

\appendix


\section{}
\label{sec:appendix}

In this section, we provide the mathematical specifications about $\alpha$-dynamic querying. Before we can give the explicit definitions of the query fractions for the next iteration, we have to derive some general restrictions on the values of the fractions $\alpha_r(t)$, $\alpha_a(t)$ and $\alpha_z(t)$ for all $t \in \{0, \dots, T\}$, where $T$ denotes the maximum number of iterations. First, a fundamental requirement of the fractions is that
\begin{equation}
\alpha_r(t) + \alpha_a(t) + \alpha_z(t) = 1,
\label{eq:unitequality}
\end{equation}
and $\alpha_r(t), \alpha_a(t), \alpha_z(t) \in [0,1]$. Furthermore, each iteration, we want to query at least one anomaly and one uncertainty, otherwise~\eqref{eq:anomdelta} or~\eqref{eq:uncertdelta} is undefined, and consequently, $\Delta^{\gamma}(t)$ is undefined. Thus, we need $\alpha_a(t)$ and $\alpha_z(t)$ to be at least $\alpha^{\min}_{a,z} := 1/Q$. Then, the number of anomalies and uncertainties in $\mathcal{Q}(t)$ is at least $Q \cdot \alpha^{\min}_{a,z} = 1$. This means the upper bounds for $\alpha_a(t)$ and $\alpha_z(t)$ are restricted to be at most $1 - \alpha^{\min}_{a,z}$. The upper bound for $\alpha_r(t)$ is at most $\alpha^{\max}_r := 1 - 2\alpha^{\min}_{a,z}$, since we do allow to query no random observations, and hence, allow $\alpha_r(t)$ to go to 0. The definition of $\alpha_r(t)$ is given in~\eqref{eq:prealpharand}. Note that $\alpha_r(t) \in [\alpha^{\min}_r, \alpha^{\max}_r) \subset [0,1]$ with $\alpha^{\min}_r :=\alpha_r(T)$. 
Now, we can define the upper bound for $\alpha_a(t)$ and $\alpha_z(t)$ as
\begin{equation*}
\alpha^{\max}_{a,z}(t) := 1 - \alpha_r(t) - \alpha^{\min}_{a,z}.
\end{equation*}
Note that this upper bound depends on the iteration number $t$.

By using \textit{(i)} $\alpha_r(t), \alpha_a(t), \alpha_z(t) \in [0,1]$, \textit{(ii)}~\eqref{eq:unitequality}, and \textit{(iii)} the definitions of $\alpha^{\min}_{a,z}$, $\alpha^{\min}_r$, $\alpha^{\max}_{a,z}(t)$ and $\alpha^{\max}_r$, we characterize the $w$ variables introduced in update rules~\eqref{eq:prealphaanom} and~\eqref{eq:prealphauncert} as follows:
\begin{align*}
w^{(1)}_a(t) &= \alpha^{\max}_{a,z}(t) - \alpha_a(t) \\
w^{(2)}_a(t) &= \alpha_a(t) - \alpha^{\min}_{a,z} \\
w^{(1)}_z(t) &= \alpha^{\max}_{a,z}(t) - \alpha_z(t) \\
w^{(2)}_z(t) &= \alpha_z(t) - \alpha^{\min}_{a,z}.
\end{align*}
Let us explain the specifics of the update of the anomaly fraction $\alpha_a(\cdot)$. We take the old value of $\alpha_a(t)$ as a starting point for $\alpha_a(t+1)$. This fraction can be increased by at most $w^{(1)}_a(t) = \alpha^{\max}_{a,z}(t) - \alpha_a(t)$ to obtain $\alpha^{\max}_{a,z}(t)$ as the new value. This increase $w^{(1)}_a(t)$ is scaled down by $\max\{0,\Delta^{\gamma}(t)\}$ such that the increment is proportional to the value of $\Delta^{\gamma}(t)$. Similarly, $\alpha_a(t)$ can be decreased by at most $w^{(2)}_a(t) = \alpha_a(t) - \alpha^{\min}_{a,z}$ to obtain $\alpha^{\min}_{a,z}$. The decrease $w^{(2)}_a(t)$ is then scaled down by $\min\{0,\Delta^{\gamma}(t)\}$. Hence, the constants ensure that $\alpha_a(t+1)$ lies in the interval $I_{a,z}(t) := [\alpha^{\min}_{a,z}, \alpha^{\max}_{a,z}(t)]$. However, it should lie in the interval $I_{a,z}(t+1) := [\alpha^{\min}_{a,z}, \alpha^{\max}_{a,z}(t+1)]$. This is why we apply the linear transformation $\lambda_{t+1}: I_{a,z}(t) \rightarrow I_{a,z}(t+1)$. This function is given by
\begin{equation}
\label{eq:linearscaling}
\lambda_{t+1}(\alpha) = \frac{\alpha^{\max}_{a,z}(t+1) - \alpha^{\min}_{a,z}}{\alpha^{\max}_{a,z}(t) - \alpha^{\min}_{a,z}}(\alpha - \alpha^{\min}_{a,z}) + \alpha^{\min}_{a,z}.
\end{equation}
Note that this function is not well-defined whenever $\alpha^{\max}_{a,z}(t) - \alpha^{\min}_{a,z} = 0$. This happens when $\alpha_r(t) = 1 - \alpha^{\min}_{a,z} = \alpha^{\max}_r$. However, the definition of $\alpha_r(t)$ in Equation~\eqref{eq:prealpharand} shows that $\alpha_r(t)$ is strictly less than $\alpha^{\max}_r$, and hence the denominator in Equation~\eqref{eq:linearscaling} cannot be zero and $\lambda_{t+1}$ is well-defined. 

Finally, for $t \in \{0, \dots, T-1\}$, the systems of equations for the three query fractions are given by
\begin{equation*}
\begin{cases}
\alpha_r(0) =  \alpha^{\max}_r \cdot 2^{-\tau \cdot L(0)} \\
\alpha_r(t+1) =  \alpha^{\max}_r \cdot 2^{-\tau \cdot (L(0) + Q \cdot (t+1))},
\end{cases} 
\end{equation*}
\begin{equation*}
\begin{cases}
\alpha_a(0) = \alpha^{(0)}_a \\
\alpha_a(t+1) =\lambda_{t+1}\Big(\alpha_a(t) &\hspace{-3mm}+ (\alpha^{\max}_{a,z}(t) - \alpha_a(t))\max\{0, \Delta^{\gamma}(t)\} \\
 &\hspace{-3mm}+ (\alpha_a(t) - \alpha^{\min}_{a,z}) \min\{0, \Delta^{\gamma}(t)\}\Big),
\end{cases}
\end{equation*}
and
\begin{equation*}
\begin{cases}
\alpha_z(0) = \alpha^{(0)}_z \\
\alpha_z(t+1) = \lambda_{t+1}\Big(\alpha_z(t) &\hspace{-3mm}+ (\alpha^{\max}_{a,z}(t) - \alpha_z(t))\max\{0, -\Delta^{\gamma}(t)\} \\
 &\hspace{-3mm}+ (\alpha_z(t) - \alpha^{\min}_{a,z}) \min\{0, -\Delta^{\gamma}(t)\}\Big).
\end{cases} 
\end{equation*}

\bibliographystyle{elsarticle-num} 
\bibliography{jas-references}

\end{document}